\documentclass[letterpaper,pra,twocolumn,floatfix,showpacs]{revtex4}

\usepackage{amssymb,amsmath,amsthm,graphicx,xspace}
\begin{document}

\newcommand{\grayed}{\color[gray]{0.3}}	

\def\Id{1\!\mathrm{l}}
\newcommand{\Tr}{\mathrm{Tr}}
\newcommand{\ket}[1]{\ensuremath{\left|\scriptstyle#1\right\rangle}}
\newcommand{\bra}[1]{\ensuremath{\left\langle\scriptstyle#1\right|}}
\newcommand{\braket}[2]{\ensuremath{\left\langle\scriptstyle#1|#2\right\rangle}}
\newcommand{\ketbra}[2]{\ket{#1}\!\!\bra{#2}}
\newcommand{\expect}[1]{\ensuremath{\left\langle#1\right\rangle}}
\newcommand{\braopket}[3]{\ensuremath{\bra{#1}#2\ket{#3}}}
\newcommand{\proj}[1]{\ketbra{#1}{#1}}
\newcommand{\oper}[1]{\ensuremath{\hat{#1}}}
\newcommand{\superop}[1]{\Hat{\Hat{#1}}}
\newcommand{\avg}[1]{\ensuremath{\overline{#1}}}
\newcommand{\Proj}[1]{\proj{#1#1}}
\newcommand{\PProj}[1]{\proj{#1#1#1#1}}
\newcommand{\NProj}[2]{\proj{{#1}^{\tensor #2}}}

\newcommand{\Oket}[1]{\ensuremath{\left|#1\right\rangle}}
\newcommand{\Obra}[1]{\ensuremath{\left\langle#1\right|}}
\newcommand{\Obraket}[2]{\ensuremath{\left\langle#1|#2\right\rangle}}
\newcommand{\Oketbra}[2]{\Sket{#1}\!\!\Sbra{#2}}
\newcommand{\Oproj}[1]{\Sketbra{#1}{#1}}
\newcommand{\Sproj}[1]{\Oproj{\proj{#1}}}

\newcommand{\Psymm}[1]{\ensuremath{\Pi^{(#1)}_{\text{symm}}}}
\newcommand{\Ens}{\ensuremath{\mathcal{E}}}
\newcommand{\Eu}{\ensuremath{\E_{\text{Haar}}}}
\renewcommand{\P}{\ensuremath{\mathcal{P}}}
\renewcommand{\S}{\ensuremath{\superop{S}}}
\newcommand{\ssd}{\ensuremath{\S\S^{\dagger}}}

\providecommand{\tensor}{\ensuremath{otimes}}
\renewcommand{\tensor}{\ensuremath{\otimes}}
\newcommand{\Tensor}{\ensuremath{\bigotimes}}
\newcommand{\diff}{\ensuremath{\mathrm{d}}}
\newcommand{\pdiff}[2]{\ensuremath{\frac{\partial #1}{\partial #2}}}

\newcommand{\z}{\ensuremath{\vec{z}}}

\newcommand{\A}{\ensuremath{\mathcal{A}}}
\newcommand{\B}{\ensuremath{\mathcal{B}}}
\newcommand{\AB}{\ensuremath{\mathcal{AB}}}

\newcommand{\R}{\ensuremath{\mathbf{R}}}
\newcommand{\Rd}{\ensuremath{\R_{\mathrm{diag}}}}
\newcommand{\Rod}{\ensuremath{\R_{\mathrm{offdiag}}}}
\newcommand{\V}{\ensuremath{\mathbf{V}}}
\newcommand{\M}{\ensuremath{\mathbf{M}}}
\newcommand{\J}{\ensuremath{\mathbf{J}}}
\newcommand{\Jx}{\ensuremath{\J_x}}
\newcommand{\Jy}{\ensuremath{\J_y}}
\newcommand{\Jz}{\ensuremath{\J_z}}

\newcommand{\Z}{\ensuremath{\mathbf{Z}}}
\newcommand{\X}{\ensuremath{\mathbf{X}}}
\newcommand{\Y}{\ensuremath{\mathbf{Y}}}
\newcommand{\Zsys}{\ensuremath{\Z_{\Sys}}}
\newcommand{\Xsys}{\ensuremath{\X_{\Sys}}}
\newcommand{\Ysys}{\ensuremath{\Y_{\Sys}}}
\newcommand{\Zenv}{\ensuremath{\Z_{\Env}}}
\newcommand{\Xenv}{\ensuremath{\X_{\Env}}}
\newcommand{\Yenv}{\ensuremath{\Y_{\Env}}}
\newcommand{\Esys}{\ensuremath{E_0}}
\newcommand{\Intstrength}{\ensuremath{g}}

\newcommand{\Sys}{\ensuremath{\mathcal{S}}}
\newcommand{\Env}{\ensuremath{\mathcal{E}}}
\newcommand{\Frag}{\ensuremath{\mathcal{F}}}
\newcommand{\Envset}[1]{\ensuremath{\Env_{\left\{#1\right\}}}}
\newcommand{\rhoS}{\ensuremath{\rho_{\Sys}}}
\newcommand{\rhoE}{\ensuremath{\rho_{\Env}}}
\newcommand{\rhoSE}{\ensuremath{\rho_{\Sys\Env}}}
\newcommand{\rhoF}{\ensuremath{\rho_{\Frag}}}
\newcommand{\rhoSF}{\ensuremath{\rho_{\Sys\Frag}}}
\newcommand{\ds}{\ensuremath{D_{\Sys}}}
\newcommand{\de}{\ensuremath{D_{\Env}}}
\newcommand{\Nenv}{\ensuremath{N_{\mathrm{env}}}}
\newcommand{\idx}[1]{^{(#1)}}

\newcommand{\I}{\ensuremath{\mathcal{I}}}
\newcommand{\Ise}{\ensuremath{\I_{\Sys:\Env}}}
\newcommand{\Isf}{\ensuremath{\I_{\Sys:\Frag}}}
\newcommand{\Isen}[1]{\ensuremath{\I_{\Sys:\Env_{#1}}}}
\newcommand{\Ibar}{\ensuremath{\overline{\I}}}

\newcommand{\Hh}{\ensuremath{H}}
\newcommand{\Hs}{\ensuremath{\Hh_{\Sys}}}
\newcommand{\Hmax}{\ensuremath{\Hh_{\mathrm max}}}
\newcommand{\He}[1][]{\ensuremath{\Hh_{\Env_{#1}}}}
\newcommand{\Hse}[1][]{\ensuremath{\Hh_{\Sys\Env_{#1}}}}

\newcommand{\dH}{\ensuremath{\delta\Hh}}
\newcommand{\Hsquig}{\ensuremath{\tilde{\Hh}}}
\newcommand{\dHsquig}{\ensuremath{\delta\Hsquig}}
\newcommand{\spr}{\ensuremath{\circ}}

\newcommand{\Ham}{\ensuremath{\mathbf{{H}}}}
\newcommand{\Hspace}{\ensuremath{\mathcal{H}}}
\newcommand{\HSspace}{\ensuremath{\mathcal{B}(\Hspace)}}

\newcommand{\Hint}{\ensuremath{\Ham_{\mathrm{int}}}}
\newcommand{\Henv}{\ensuremath{\Ham_{\mathrm{env}}}}
\newcommand{\Hsys}{\ensuremath{\Ham_{\mathrm{sys}}}}
\newcommand{\Hee}{\ensuremath{\Ham_{\Env-\Env}}}
\newcommand{\Sint}{\ensuremath{\mathcal{O}_{\Sys}}}
\newcommand{\Eint}{\ensuremath{\mathcal{O}_{\Env}}}

\newcommand{\fI}{\ensuremath{f_{\I}}}
\newcommand{\fcap}{\ensuremath{f_{cap}}}
\newcommand{\hf}{\ensuremath{\frac12}}
\newcommand{\SU}[1]{\ensuremath{SU\left(#1\right)}}
\newcommand{\psuff}{\ensuremath{p_{\mathrm{suff}}}}
\newcommand{\preq}{\ensuremath{p_{\mathrm{req}}}}
\newcommand{\erf}{\ensuremath{\mathrm{erf}}}

\newcommand{\CP}[1]{\ensuremath{\mathbb{C}P^{#1}}}
\newcommand{\RP}[1]{\ensuremath{\mathbb{R}P^{#1}}}

\newcommand{\Msys}{\ensuremath{m_{\Sys}}}
\newcommand{\Menv}[1]{\ensuremath{m_{#1}}}
\newcommand{\omegasys}{\ensuremath{\omega_{\Sys}}}
\newcommand{\omegaenv}[1]{\ensuremath{\omega_{#1}}}
\newcommand{\omegabare}{\ensuremath{\Omega_0}}
\newcommand{\omegashift}{\ensuremath{\delta\Omega}}
\newcommand{\CC}[1]{\ensuremath{C_{#1}}}
\newcommand{\xsys}{\ensuremath{x_{\Sys}}}
\newcommand{\xenv}[1]{\ensuremath{y_{#1}}}
\newcommand{\psys}{\ensuremath{p_{\Sys}}}
\newcommand{\penv}[1]{\ensuremath{q_{#1}}}
\newcommand{\Var}{\ensuremath{\mathcal{V}}}
\newcommand{\T}{\ensuremath{\mathcal{T}}}
\newcommand{\commutator}[2]{\ensuremath{\left[#1,#2\right]}}
\newcommand{\anticommutator}[2]{\ensuremath{\left\{#1,#2\right\}}}
\newcommand{\omeff}{\ensuremath{\omega_{\mathrm{eff}}}}
\newcommand{\gameff}{\ensuremath{\gamma_{\mathrm{eff}}}}
\newcommand{\Heff}{\ensuremath{\Ham_{\mathrm{eff}}}}
\newcommand{\Inr}{\ensuremath{\I_{\mathrm{NR}}}}
\newcommand{\Ir}{\ensuremath{\I_{\mathrm{R}}}}
\newcommand{\Iq}{\ensuremath{\I_{\mathrm{Q}}}}
\newcommand{\EMConst}{\ensuremath{\gamma_{\scriptscriptstyle{\mathrm{EM}}}}}
\newcommand{\chunk}{fragment\xspace}
\newcommand{\C}{\ensuremath{\mathcal{C}}}

\newcommand{\xbar}{\ensuremath{\bar{x}}}
\newcommand{\ybar}{\ensuremath{\bar{y}}}
\newcommand{\xp}{\ensuremath{x^+}}
\newcommand{\xm}{\ensuremath{x^-}}
\newcommand{\yp}{\ensuremath{y^+}}
\newcommand{\ym}{\ensuremath{y^-}}
\newcommand{\xo}{\ensuremath{x_0}}
\newcommand{\yo}{\ensuremath{y_0}}
\newcommand{\xobar}{\ensuremath{\bar{x_0}}}
\newcommand{\yobar}{\ensuremath{\bar{y_0}}}
\newcommand{\xpo}{\ensuremath{x_0^+}}
\newcommand{\xmo}{\ensuremath{x_0^-}}
\newcommand{\ypo}{\ensuremath{y_0^+}}
\newcommand{\ymo}{\ensuremath{y_0^-}}
\newcommand{\Xbar}{\ensuremath{\bar{X}}}
\newcommand{\Ybar}{\ensuremath{\bar{Y}}}
\newcommand{\Xpls}{\ensuremath{X^+}}
\newcommand{\Xmns}{\ensuremath{X^-}}
\newcommand{\Ypls}{\ensuremath{Y^+}}
\newcommand{\Ymns}{\ensuremath{Y^-}}
\newcommand{\Xo}{\ensuremath{X_0}}
\newcommand{\Yo}{\ensuremath{Y_0}}
\newcommand{\Xobar}{\ensuremath{\bar{X_0}}}
\newcommand{\Yobar}{\ensuremath{\bar{Y_0}}}
\newcommand{\Xopls}{\ensuremath{X_0^+}}
\newcommand{\Xomns}{\ensuremath{X_0^-}}
\newcommand{\Yopls}{\ensuremath{Y_0^+}}
\newcommand{\Yomns}{\ensuremath{Y_0^-}}
\newcommand{\Po}{\ensuremath{P_0}}
\newcommand{\Qo}{\ensuremath{Q_0}}
\newcommand{\po}{\ensuremath{p_0}}
\newcommand{\qo}{\ensuremath{q_0}}

\newcommand{\Mq}{\ensuremath{\mathcal{M}}}
\newcommand{\Rq}{\ensuremath{\mathcal{R}}}
\newcommand{\Sm}{\ensuremath{\mathcal{S}}}
\newcommand{\U}{\ensuremath{\mathcal{U}}}
\newcommand{\Q}{\ensuremath{\mathcal{Q}}}
\newcommand{\Qpm}{\ensuremath{\mathcal{Q}_{\pm}}}
\newcommand{\dpls}{\ensuremath{\partial_+}}
\newcommand{\dmns}{\ensuremath{\partial_-}}
\newcommand{\zpm}{\ensuremath{\vec{z}_{\pm}}}
\newcommand{\ct}{\ensuremath{\cos\theta}}
\newcommand{\cnt}[1]{\ensuremath{\cos^{#1}\theta}}
\newcommand{\st}{\ensuremath{\sin\theta}}
\newcommand{\snt}[1]{\ensuremath{\sin^{#1}\theta}}
\newcommand{\cwt}{\ensuremath{\cos(\omega t)}}
\newcommand{\cnwt}[1]{\ensuremath{\cos^{#1}(\omega t)}}
\newcommand{\swt}{\ensuremath{\sin(\omega t)}}
\newcommand{\snwt}[1]{\ensuremath{\sin^{#1}(\omega t)}}
\newcommand{\clt}{\ensuremath{\cos(\lambda t)}}
\newcommand{\cnlt}[1]{\ensuremath{\cos^{#1}(\lambda t)}}
\newcommand{\slt}{\ensuremath{\sin(\lambda t)}}
\newcommand{\snlt}[1]{\ensuremath{\sin^{#1}(\lambda t)}}
\newcommand{\chlt}{\ensuremath{\cosh(\lambda t)}}
\newcommand{\chnlt}[1]{\ensuremath{\cosh^{#1}(\lambda t)}}
\newcommand{\shlt}{\ensuremath{\sinh(\lambda t)}}
\newcommand{\shnlt}[1]{\ensuremath{\sinh^{#1}(\lambda t)}}
\newcommand{\cws}{\ensuremath{\cos(\omega s)}}
\newcommand{\sws}{\ensuremath{\sin(\omega s)}}
\newcommand{\cls}{\ensuremath{\cos(\lambda s)}}
\newcommand{\sls}{\ensuremath{\sin(\lambda s)}}
\newcommand{\chls}{\ensuremath{\cosh(\lambda s)}}
\newcommand{\shls}{\ensuremath{\sinh(\lambda s)}}
\newcommand{\cvec}[4]{\ensuremath{\left(\begin{array}{c}#1\\#2\\#3\\#4\end{array}\right)}}
\newcommand{\rvec}[4]{\ensuremath{\left(#1,#2,#3,#4\right)}}
\newcommand{\Ft}{\ensuremath{\hat{F}}}
\newcommand{\Denom}{\ensuremath{\begin{array}{l}\left(\omega^2-\lambda^2\right)\cnt2\snt2\swt\shlt \\
	+\omega\lambda\left(2\cwt\chlt\cnt2\snt2+\cnt4+\snt4\right)\end{array}}}

\newcommand{\syssymbol} {\ensuremath{\mathcal{S}}}
\newcommand{\envsymbol} {\ensuremath{\mathcal{E}}}
\newcommand{\HS} {\ensuremath{\Ham_{\syssymbol}}}
\newcommand{\HE} {\ensuremath{\Ham_{\envsymbol}}}
\newcommand{\HSE} {\ensuremath{\Ham_{\syssymbol\envsymbol}}}
\newcommand{\sysop} {\ensuremath{\hat{\mathcal{O}}_{\syssymbol}}}
\newcommand{\envop} {\ensuremath{\hat{\mathcal{O}}_{\envsymbol}}}
\newcommand{\sysstate}[1] {\ensuremath{\ket{\Psi_{\syssymbol}#1}}}
\newcommand{\envstate}[1] {\ensuremath{\ket{\Psi_{\envsymbol}#1}}}
\newcommand{\superstate}[1] {\ensuremath{\ket{\Psi_{\syssymbol\envsymbol}#1}}}
\newcommand{\syseigenstate}[2] {\ensuremath{\ket{s_{#1}#2}}}
\newcommand{\syseigenvalue} {\ensuremath{s}}
\newcommand{\enveigenvalue} {\ensuremath{\varepsilon}}
\newcommand{\enveigenstate}[2] {\ensuremath{\ket{\varepsilon_{#1}#2}}}
\newcommand{\varentropy} {\ensuremath{\tilde{\Hh}}}
\newcommand{\entropy} {\Hh}
\newcommand{\energy} {E}
\newcommand{\enveigenstateoverlaps} {\ensuremath{\braket{\varepsilon_n(t)}{\varepsilon_m(t)}}}
\newcommand{\Vp}{\ensuremath{\mathcal{V}_p}}
\newcommand{\Ve}{\ensuremath{\mathcal{V}_\envsymbol}}

\newtheorem{lemma}{Lemma}
\newtheorem{theorem}{Theorem}
\newtheorem{corollary}{Corollary}
\newtheorem{definition}{Definition}

\def\FCW{1.0\columnwidth}
\def\HCW{0.5\columnwidth}
\def\FTW{0.935\textwidth}
\def\HTW{0.47\textwidth}
\def\TTW{0.33\textwidth}

\def\figstart{\begin{figure}}
\def\figend{\end{figure}}
\def\wfigstart{\begin{figure*}}
\def\wfigend{\end{figure*}}

\def\widetextstart{\begin{widetext}}
\def\widetextend{\end{widetext}}

\newcommand{\OnePanelFigure}[4][One plot]{\figstart[tb!]\includegraphics[width=\FCW]{#3}\caption[#1]{#4}\label{#2}\figend}

\newcommand{\OnePanelFigureNow}[4][One plot]{\figstart[tbp!]\includegraphics[width=\FCW]{#3}\caption[#1]{#4}\label{#2}\figend}

\newcommand{\OneLargePanelFigure}[4][One plot]{\wfigstart[tbp!]\includegraphics[width=\FTW]{#3}\caption[#1]{#4}\label{#2}\wfigend}

\newcommand{\OneSmallPanelFigure}[4][One plot]{\figstart[tb]\begin{center}\includegraphics[width=\HTW]{#3}\end{center}\caption[#1]{#4}\label{#2}\figend}

\newcommand{\TwoPanelFigure}[5][Two plots]{\wfigstart[tb]\begin{tabular}{ll}\textbf{(a)}&\textbf{(b)}\\
\includegraphics[width=\HTW]{#3}&\includegraphics[width=\HTW]{#4}\\ \end{tabular}\caption[#1]{#5}\label{#2}\wfigend}

\newcommand{\TwoSmallPanelFigure}[5][Two plots]{\figstart[tb]\begin{tabular}{ll}\textbf{(a)}&\textbf{(b)}\\
\includegraphics[width=\HCW]{#3}&\includegraphics[width=\HCW]{#4}\\ \end{tabular}\caption[#1]{#5}\label{#2}\figend}

\newcommand{\TwoPanelFigureNow}[5][Two plots]{\wfigstart[tb!]\begin{tabular}{ll}\textbf{(a)}&\textbf{(b)}\\
\includegraphics[width=\HTW]{#3}&\includegraphics[width=\HTW]{#4}\\ \end{tabular}\caption[#1]{#5}\label{#2}\wfigend}

\newcommand{\ThreePanelFigure}[6][Three plots]{\wfigstart[tbp!]\begin{tabular}{lll}
\textbf{(a)}&\textbf{(b)}&\textbf{(c)} \\
\includegraphics[width=\TTW]{#3}& \includegraphics[width=\TTW]{#4}& \includegraphics[width=\TTW]{#5}
\end{tabular} \caption[#1]{#6}\label{#2}\wfigend}

\newcommand{\FourPanelFigure}[7][Four plots]{\wfigstart[tb!]\begin{tabular}{ll}\textbf{(a)}&\textbf{(b)}\\
\includegraphics[width=\HTW]{#3}&\includegraphics[width=\HTW]{#4}\\ \textbf{(c)}&\textbf{(d)}\\
\includegraphics[width=\HTW]{#5}&\includegraphics[width=\HTW]{#6}\\ \end{tabular}\caption[#1]{#7}\label{#2}\wfigend}

\newcommand{\FourSmallPanelFigure}[7][Four plots]{\figstart[tb!]\begin{tabular}{ll}\textbf{(a)}&\textbf{(b)}\\
\includegraphics[width=\HCW]{#3}&\includegraphics[width=\HCW]{#4}\\ \textbf{(c)}&\textbf{(d)}\\
\includegraphics[width=\HCW]{#5}&\includegraphics[width=\HCW]{#6}\\ \end{tabular}\caption[#1]{#7}\label{#2}\figend}

\newcommand{\SixPanelFigure}[9][Six plots]{\wfigstart[tbp!]\begin{tabular}{ll}
\textbf{(a)}&\textbf{(b)}\\ \includegraphics[width=\HTW]{#3}&\includegraphics[width=\HTW]{#4}\\
\textbf{(c)}&\textbf{(d)}\\ \includegraphics[width=\HTW]{#5}&\includegraphics[width=\HTW]{#6}\\
\textbf{(e)}&\textbf{(f)}\\ \includegraphics[width=\HTW]{#7}&\includegraphics[width=\HTW]{#8}\\
\end{tabular} \caption[#1]{#9}\label{#2}\wfigend}

\newcommand{\SixSmallPanelFigure}[9][Six plots]{\figstart[tbp!]\begin{tabular}{ll}
\textbf{(a)}&\textbf{(b)}\\ \includegraphics[width=\HCW]{#3}&\includegraphics[width=\HCW]{#4}\\
\textbf{(c)}&\textbf{(d)}\\ \includegraphics[width=\HCW]{#5}&\includegraphics[width=\HCW]{#6}\\
\textbf{(e)}&\textbf{(f)}\\ \includegraphics[width=\HCW]{#7}&\includegraphics[width=\HCW]{#8}\\
\end{tabular} \caption[#1]{#9}\label{#2}\figend}

\newcommand{\SixPanelFigureCondensed}[9][Six plots]{\wfigstart[tbp!]\begin{tabular}{lll}
\textbf{(a)}&\textbf{(c)}&\textbf{(e)} \\
\includegraphics[width=\TTW]{#3}& \includegraphics[width=\TTW]{#5}& \includegraphics[width=\TTW]{#7} \\ 
\textbf{(b)}&\textbf{(d)}&\textbf{(f)} \\
\includegraphics[width=\TTW]{#4}& \includegraphics[width=\TTW]{#6}& \includegraphics[width=\TTW]{#8} \\
\end{tabular} \caption[#1]{#9}\label{#2}\wfigend}

\newcommand{\TwoPanelFigureX}[5][Two plots]{\wfigstart[tb]\resizebox{\FTW}{!}{\textbf{(a)}
\includegraphics{#3}\textbf{(b)}\includegraphics{#4}}\caption[#1]{#5}\label{#2}\wfigend}

\newcommand{\FourPanelFigureX}[7][Four plots]{\wfigstart[tb!]\resizebox{\FTW}{!}{\begin{tabular}{ll}
\textbf{(a)}&\textbf{(b)} \\ \includegraphics{#3}& \includegraphics{#4} \\ 
\textbf{(c)}&\textbf{(d)} \\ \includegraphics{#5}& \includegraphics{#6} \\
\end{tabular}} \caption[#1]{#7}\label{#2}\wfigend}

\newcommand{\FourSmallPanelFigureX}[7][Four plots]{\figstart[tb!]\resizebox{\FCW}{!}{\begin{tabular}{ll}
\textbf{(a)}&\textbf{(b)} \\ \includegraphics{#3}& \includegraphics{#4} \\ 
\textbf{(c)}&\textbf{(d)} \\ \includegraphics{#5}& \includegraphics{#6} \\
\end{tabular}} \caption[#1]{#7}\label{#2}\figend}

\newcommand{\SixPanelFigureX}[9][Six plots]{\wfigstart[tb!]\resizebox{\FTW}{!}{\begin{tabular}{ll}
\textbf{(a)}&\textbf{(b)} \\ \includegraphics{#3}& \includegraphics{#4} \\ 
\textbf{(c)}&\textbf{(d)} \\ \includegraphics{#5}& \includegraphics{#6} \\
\textbf{(e)}&\textbf{(f)} \\ \includegraphics{#7}& \includegraphics{#8} \\
\end{tabular}} \caption[#1]{#9}\label{#2}\wfigend}

\newcommand{\SixSmallPanelFigureX}[9][Six plots]{\figstart[tb!]\resizebox{\FCW}{!}{\begin{tabular}{ll}
\textbf{(a)}&\textbf{(b)} \\ \includegraphics{#3}& \includegraphics{#4} \\ 
\textbf{(c)}&\textbf{(d)} \\ \includegraphics{#5}& \includegraphics{#6} \\
\textbf{(e)}&\textbf{(f)} \\ \includegraphics{#7}& \includegraphics{#8} \\
\end{tabular}} \caption[#1]{#9}\label{#2}\figend}

\newcommand{\SixPanelFigureCondensedX}[9][Six plots]{\wfigstart[tb!]\resizebox{\FTW}{!}{\begin{tabular}{lll}
\textbf{(a)}&\textbf{(b)}&\textbf{(c)} \\ \includegraphics{#3}& \includegraphics{#4}& \includegraphics{#5} \\ 
\textbf{(d)}&\textbf{(e)}&\textbf{(f)} \\ \includegraphics{#6}& \includegraphics{#7}& \includegraphics{#8}  \\
\end{tabular}} \caption[#1]{#9}\label{#2}\wfigend}

\newcommand{\TwoLongPanelFigure}[5][Two long plots]{\wfigstart[tb!]\begin{tabular}{l}
\textbf{(a)} \\ \includegraphics[width=\FTW]{#3} \\ \textbf{(b)} \\ \includegraphics[width=\FTW]{#4}
\end{tabular}\caption[#1]{#5}\label{#2}\wfigend}

\newcommand{\TwoSmallLongPanelFigure}[5][Two long plots]{\figstart[tb!]\begin{tabular}{l}
\textbf{(a)} \\ \includegraphics[width=\FCW]{#3} \\ \textbf{(b)} \\ \includegraphics[width=\FCW]{#4}
\end{tabular}\caption[#1]{#5}\label{#2}\figend}

\newcommand{\TwoLongSplitPanelFigure}[7][Two split plots]{\wfigstart[tb!]\resizebox{\FTW}{!}{\textbf{(a)} \includegraphics{#3}\includegraphics{#4}}\\ \resizebox{\FTW}{!}{\textbf{(b)}\includegraphics{#5}\includegraphics{#6}}\caption[#1]{#7}\label{#2}\wfigend}

\newcommand{\ThreeLongPanelFigure}[6][Three long plots]{\wfigstart[tb!]\begin{tabular}{l}
\textbf{(a)} \\ \includegraphics[width=\FTW]{#3} \\
\textbf{(b)} \\ \includegraphics[width=\FTW]{#4} \\ 
\textbf{(c)} \\ \includegraphics[width=\FTW]{#5}
\end{tabular}\caption[#1]{#6}\label{#2}\wfigend}

\newcommand{\ThreeSmallLongPanelFigure}[6][Three long plots]{\figstart[tb!]\begin{tabular}{l}
\textbf{(a)} \\ \includegraphics[width=\FCW]{#3} \\
\textbf{(b)} \\ \includegraphics[width=\FCW]{#4} \\ 
\textbf{(c)} \\ \includegraphics[width=\FCW]{#5}
\end{tabular}\caption[#1]{#6}\label{#2}\figend}

\newcommand{\FourLongPanelFigure}[7][Four long plots]{\wfigstart[tbp!]\begin{tabular}{l}
\textbf{(a)} \\ \includegraphics[width=\FTW]{#3} \\ \textbf{(b)} \\ \includegraphics[width=\FTW]{#4} \\
\textbf{(c)} \\ \includegraphics[width=\FTW]{#5} \\ \textbf{(d)} \\ \includegraphics[width=\FTW]{#6}
\end{tabular}\caption[#1]{#7}\label{#2}\wfigend}

\newcommand{\FourSmallLongPanelFigure}[7][Four long plots]{\figstart[tbp!]\begin{tabular}{l}
\textbf{(a)} \\ \includegraphics[width=\FCW]{#3} \\ \textbf{(b)} \\ \includegraphics[width=\FCW]{#4} \\
\textbf{(c)} \\ \includegraphics[width=\FCW]{#5} \\ \textbf{(d)} \\ \includegraphics[width=\FCW]{#6}
\end{tabular}\caption[#1]{#7}\label{#2}\figend}

\ifx\pdftexversion\undefined
  \newcommand{\grext}{eps}
\else
  \newcommand{\grext}{pdf}
\fi
\DeclareGraphicsRule{*}{\grext}{*}{}

\newcommand{\paperchapter}{paper\xspace}
\newcommand{\staticsref}{\cite{RBK05b}\xspace}
\newcommand{\dynamicsref}{\cite{RBK05c}\xspace}
\newcommand{\ptralgref}{\cite{RBK05d}\xspace}
\newcommand{\qbmref}{\cite{RBK05e}\xspace}
\newcommand{\connectionsref}{\cite{RBK05e}\xspace}
\newcommand{\amplificationref}{\cite{RBKPRA03}\xspace}
\newcommand{\startwideequation}{\begin{widetext}}
\newcommand{\stopwideequation}{\end{widetext}}

\graphicspath{{Figures/}}

\title{Quantum Darwinism: Entanglement, Branches, and the Emergent Classicality of Redundantly Stored Quantum Information}
\date{\today}
\author{Robin Blume-Kohout}
\email{robin@blumekohout.com}
\affiliation{Theoretical Division, Los Alamos National Laboratory, Los Alamos, NM 87545}
\affiliation{Institute for Quantum Information, California Institute of Technology, Pasadena, CA 91125}
\author{Wojciech H. Zurek}
\email{whz@lanl.gov}
\affiliation{Theoretical Division, Los Alamos National Laboratory, Los Alamos, NM 87545}
\begin{abstract}
We lay a comprehensive foundation for the study of redundant information storage in decoherence processes.  Redundancy has been proposed as a prerequisite for \emph{objectivity}, the defining property of classical objects.  We consider two ensembles of states for a model universe consisting of one system and many environments: the first consisting of arbitrary states, and the second consisting of ``singly-branching'' states consistent with a simple decoherence model.  Typical states from the random ensemble do not store information about the system redundantly, but information stored in branching states has a redundancy proportional to the environment's size.  We compute the specific redundancy for a wide range of model universes, and fit the results to a simple first-principles theory.  Our results show that the presence of redundancy divides information about the system into three parts: classical (redundant); purely quantum; and the borderline, undifferentiated or ``nonredundant,'' information.
\end{abstract}


\maketitle

The theory of decoherence \cite{ZurekRMP03,SchlosshauerRMP04,PazLH01,Giulini96} has resolved much of the decades-old confusion about the transition from quantum to classical physics (see articles in \cite{WheelerBook83}).  It provides a mechanism -- weak measurement by the environment -- by which a quantum system can be compelled to behave classically.  The recent development of quantum information theory has encouraged an information-theoretic view of decoherence, wherein information about a central system ``leaks out'' into the environment, and thereby becomes classical \cite{ZurekADP00}.

In this paper, we pursue a natural extension of the decoherence program, by asking ``What happens to the information that leaks out of the system?''  That information should be sought in the ``rest of the universe'' -- i.e., the system's environment.  The environment is a witness to the system's state, and can serve as a resource for measuring or controlling the system.  Our particular focus, within this \emph{Environment as a Witness} paradigm, is on how \emph{redundantly} information about the system is recorded in the environment.  This is relevant to quantum technology; a detailed picture of how decoherence destroys quantum information may help in designing schemes to correct its effects.

It also illuminates fundamental physics.  Massive redundancy can cause certain information to become \emph{objective}, at the expense of other information.  The process by which this ``fittest'' information is propagated through the environment, at the expense of incompatible information, is \emph{Quantum Darwinism}.  Two forthcoming papers (\cite{RBK05c,RBK05e}) will investigate the \emph{dynamics} of quantum Darwinism.

This paper is focused on the \emph{kinematics} of information storage and the environment-as-a-witness paradigm.  It is organized as follows.  In Section \ref{secDefiningRedundancy}, we introduce objectivity and the ``environment as a witness'' paradigm, show that redundant records indicate objectivity, and propose quantitative and qualitative measures of redundancy.  In Section \ref{secUniformEnsemble}, we analyze randomly distributed states, show that they do not display redundant information storage, and argue that they do not describe the Universe (see next paragraph) in which we live.  In Section \ref{secBranchingStates}, we propose \emph{singly-branching} states as an alternative description, and use numerics to demonstrate redundant information storage.  Section \ref{secBranchingStateTheory} presents an analytical model for the numerical results.  Finally, we summarize our most important results and discuss future work in Section \ref{secStaticsConclusions}.

We use the word ``universe'' to denote both (a) everything that exists in reality, and (b) a self-contained model of a system and its environment.  We distinguish the two by capitalizing usage (a).  Thus, while living in the \emph{Universe}, we simulate assorted \emph{universes}.

\section{The Environment as a Witness} \label{secDefiningRedundancy}

Previous studies of decoherence have focused on the system's reduced density matrix ($\rhoS$), and on master equations that describe its evolution.  To study information flow into the environment, we require a new paradigm.

We begin with a simple observation:  \textbf{information about a system ($\Sys$) is obtained by measuring its environment ($\Env$)} (see \cite{ZurekRMP03, ZurekPTP93}).  Although the standard
theories of quantum measurement (see e.g. Von Neumann \cite{VonNeumannBook55}, etc.) presume a direct measurement on the system, real experiments rely on \emph{indirect} measurements.  As you read this \paperchapter, you measure the albedo of the page -- but actually, your eyes are capturing photons from the electromagnetic environment.  Information about the page is inferred from assumed correlations between text and photons.  A similar argument holds for every physics experiment; the scientist gets information about $\Sys$ by capturing and measuring a \emph{fragment} of $\Env$.

This motivates us to focus on correlations between $\Sys$ and individual fragments of $\Env$.  In particular, we will seek to determine whether a particular state -- or a particular ensemble of states -- allows an observer who captures a small fragment of $\Env$ to deduce the system's state.  If so, then the system's state is \emph{objectively} recorded.

\subsection{Objectivity}

A property -- e.g., the state of a system -- is objective when many independent observers agree about it.  The observers' independence is crucial.  When many secondary observers are informed by a single primary observer, then only the primary observer's opinion is objective, \emph{not} necessarily the property which he observed.
\emph{Independent} observers, examining a single quantum system, cannot have agreed on a particular measurement basis beforehand.  They will generally measure different observables -- and therefore will not agree afterward.  An isolated quantum system's state cannot be objective, because measurements of noncommuting observables invalidate each other. 

Classical theory, on the other hand, permits observers to measure a system without disturbing it.  Properties of classical systems (e.g., classical states) are thus objective.  Each observer can record the state in question without altering it, and afterward all the observers will agree on what they discovered.  Of course, observers may obtain \emph{different} information -- e.g., one observer may make a more effective measurement than another -- but not \emph{contradictory} information.


Objectivity provides an excellent criterion for exploring the emergence of classicality through decoherence.  A quantum system becomes more classical as its measurable properties become more objective.  The use of ``measurable'' is significant.  Nothing can make \emph{every} property of a quantum system objective, because some observables are incompatible with others. Two observers can never simultaneously obtain reliable information about incompatible observables (such as position and momentum) of the same system.  Decoherence partially solves this problem by destroying all the observables incompatible with a system's \emph{pointer observable}.  We are thus motivated to explore (a) how the pointer observable becomes objective, and (b) how decoherence and the emergence of objectivity are related.

\subsection{Technical details and assumptions}

This ``environment as a witness'' paradigm \cite{ZurekRMP03,OllivierPRL04,Ollivier04,ZurekADP00} is ideally suited to exploring objectivity.  In order to make \emph{independent} measurements of $\Sys$, multiple observers must partition the environment into fragments.  In this paper, we assume that measurements must be made on distinct Hilbert spaces in order to be independent, so we divide the environment into fragments as
\begin{equation}
\Env = \Env_A \otimes \Env_B \otimes \Env_C \otimes \ldots.
\end{equation}

Several factors limit an observer's ability to obtain information about $\Sys$ by measuring a fragment of the environment ($\Env_A$).  We can make more or less optimistic assumptions about some of these factors, but \textbf{the degree of correlation between $\Sys$ and $\Env_A$ is clearly a limiting factor}.  An observer whose particular fragment is not correlated with $\Sys$ has no way to obtain information about $\Sys$.  That fragment of $\Env$ is irrelevant and, for the purpose of gaining information about $\Sys$, might as well not exist.  The absolute prerequisite for demonstrating a property's objectivity is that information about it be recorded in many fragments -- that is, \emph{redundantly}.

We quantify redundancy by counting the number of fragments which can provide sufficient information.  The redundancy of information about some property is a natural measure of that property's objectivity \cite{ZurekRMP03}.  Classical properties are objective because information about them is recorded with [effectively] infinite redundancy.  For instance, if we flip a coin, then its final orientation is recorded by trillions of scattered photons.  Thousands of cameras, each capturing a tiny fraction of them, could each provide a record.  Redundancy is not dependent on \emph{actual} observers.  Instead, it is a statement about what observers \emph{could} do, if they existed.

A pertinent question is ``Why not allow an observer to measure the system itself?''  First, only one observer could be allowed to do so without sacrificing independence.  Thus, at most, this would increase redundancy by 1.  Furthermore, an observer with access to the central system could measure it in some weird basis, thus destroying its state.  Since it's not then clear what the information obtained by the \emph{other} observers would refer to, we regard the system itself as off limits to observers.


\subsection{The overall program}

The work presented here is a natural extension of the decoherence program.  However, employing the environment as a communication channel -- not just ``sink'' for information lost to decoherence -- is also in a sense ``beyond decoherence.''  It is the next stage in exploring how classicality emerges from the quantum substrate. 

In order to fully understand the role that redundancy and objectivity play in (1) the emergence of classicality, and (2) the destruction of quantum coherence, we'd like to answer the following questions:
\begin{enumerate}
\item Given a state $\rhoSE$ for the system and its environment (the ``universe''), how do we quantify the redundancy of information (about $\Sys$) in $\Env$?
\item For a particular ``universe,'' what states are \emph{typical} (that is, likely to exist)?  Do they display redundancy?  If so, how much?
\item What sorts of (a) initial states, and (b) dynamics lead dynamically to redundancy?
\item Do realistic models of decoherence produce the massive redundancy we expect in the classical regime?
\item For complicated systems, with many independent properties, how do we distinguish \emph{what} property a bit of information is about?
\item When information about an observable is redundantly recorded, is information about incompatible observables inaccessible?
\end{enumerate}
The building blocks of this work -- e.g., the reasoning presented in this section -- have been laid in recent years by \cite{ZurekPTRSA98, ZurekADP00, ZurekRMP03, OllivierPRL04, Zurek03}.  The first attempt to address items (1) and (3) appeared in \cite{ZurekADP00}, and was refined in \cite{OllivierPRL04}, which also analyzed a particular simple model of decoherence numerically.  In the current \paperchapter, we answer (1) and (2) in detail, and consider (3) briefly.

\subsection{Computing redundancy} \label{secQuantitativeR}

To compute the redundancy ($R$) of some information ($\I$), we divide the environment into fragments ($\Env = \Env_A \otimes \Env_B \otimes \ldots$), and demand that each fragment supply $\I$ independently.  The redundancy of $\I$ is the number of such fragments into which the environment can be divided.  A generalized GHZ state is a good example:
\begin{equation}
\ket\psi_{\Sys\Env} = \alpha\ket{0}_{\Sys}\ket{00000\ldots 0}_{\Env} + \beta\ket{1}_{\Sys}\ket{11111\ldots 1}_{\Env} \label{eqGHZ}
\end{equation}
We can determine the system's state by measuring any sub-environment.  Each qubit in $\Env$ provides all the available information about $\Sys$ (see, however, note \footnote{We must make the \emph{right} measurement -- in this case, one which distinguishes $\ket{0}$ from $\ket{1}$ -- in order to get the information.  In this work, the amount of information that one subenvironment has is always \emph{maximized} over all possible measurements.}).  To extend this analysis to arbitrary states, we need (a) a metric for information, (b) a protocol for dividing the environment into fragments, and (c) an idea of how much of $\I$ is ``available''.

\subsubsection{A metric for information}


\OneSmallPanelFigure[Different locality (tensor product) structures for the ``universe'']{figSubdivision}{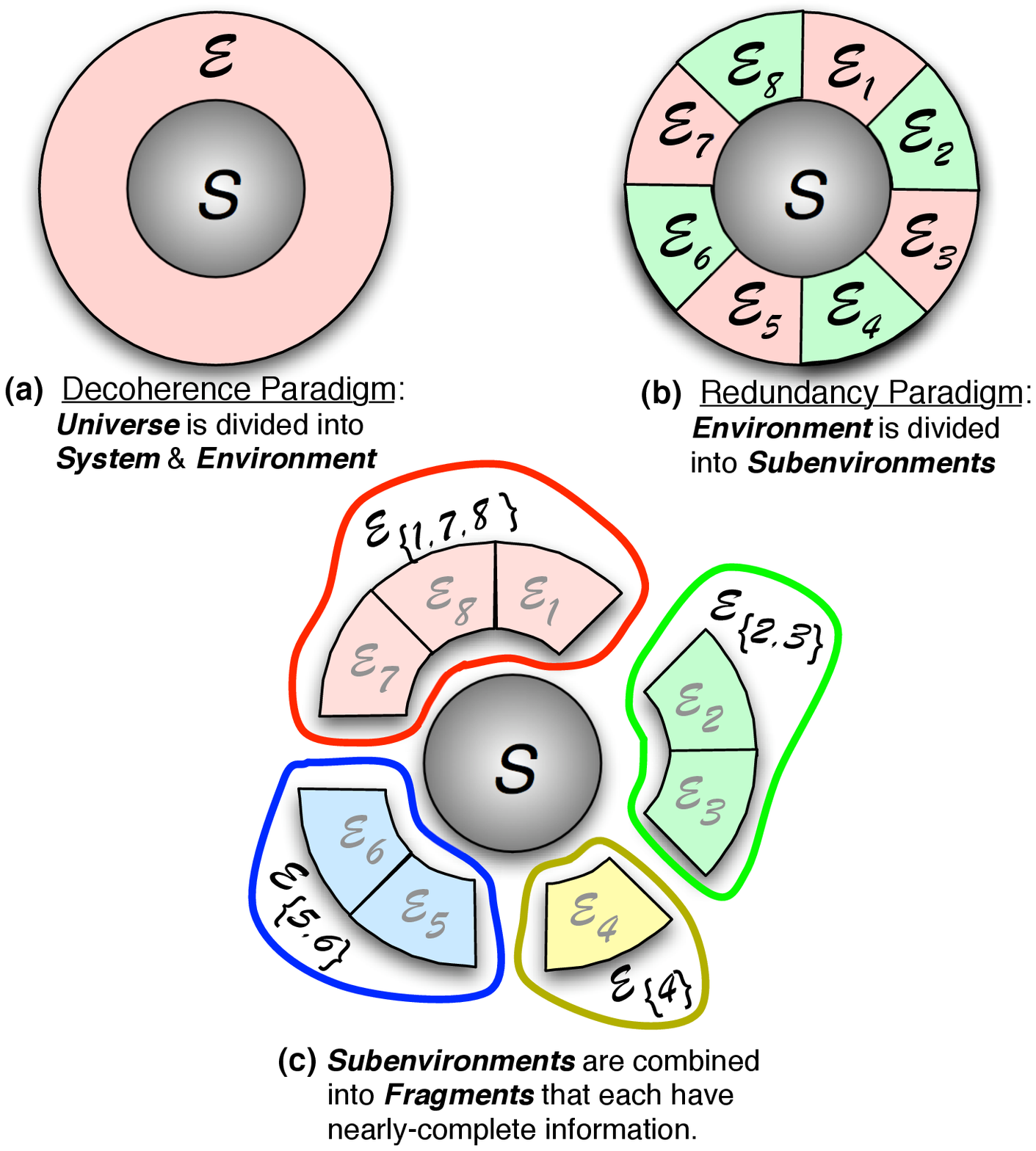}
{(Color) \textbf{Three ways to divide up the universe}.  The decoherence paradigm divides the universe into a system ($\Sys$) and an environment ($\Env$) as in \textbf{(a)}.  In the environment-as-a-witness paradigm, we further subdivide $\Env$ into \emph{subenvironments}, as in \textbf{(b)}.  No subenvironment can be further subdivided, and it is easier to measure one $\Env_i$ than to make a joint measurement on several.  \emph{Fragments} are constructed, so as to provide enough information to infer the state of $\Sys$, by combining subenvironments as in \textbf{(c)}.  Measurements on distinct fragments always commute.}

We use \emph{quantum mutual information} (QMI) as an information metric.  QMI is a generalization of the classical mutual information \cite{CoverBook91}.  \emph{Quantum} mutual information is defined in terms of the Von Neumann entropy, $H = -\Tr(\rho\log\rho)$, as:
\begin{equation}
\I_{A:B} = H_A + H_B - H_{AB}
\end{equation}
This is simple to calculate, provides a reliable measure of correlation between systems, and has been used previously for this purpose \cite{ZurekPRD82,Zurek83,ZurekRMP03}.  Unlike classical mutual information, the QMI between system $A$ and system $B$ is not bounded by the entropy of either system.  In the presence of entanglement, the QMI can be as large as $H_A + H_B$, which reflects the existence of quantum correlations beyond the classical ones \cite{OllivierPRL02}..  

\subsubsection{Dividing $\Env$ into fragments}

A pre-existing concept of locality, usually expressed as a fixed tensor product structure or as a set of allowable structures, is \emph{fundamental} to redundancy analysis.  Allowing an \emph{arbitrary} division of $\Env$ into fragments would make every state where $\Sys$ is entangled with $\Env$ (see note \footnote{In this work, we assume that the universe is in a pure state.  Any correlation between $\Sys$ and $\Env$ is due to entanglement.  Similar conclusions seem to apply when the environment is initially mixed, but we have not investigated these cases exhaustively.}) equivalent (via re-division of $\Env$) to a GHZ-like state (Eq. \ref{eqGHZ}).  Decoherence would be equivalent to redundancy.

The need for a fixed tensor product structure is familiar; both decoherence and entanglement are meaningless without a fixed division between the system and its environment (\cite{ZurekRMP03,ZurekPTP93}, see e.g. \cite{ZanardiPRL04} for a discussion of a possible 
tensor product structures' origins in measurable observables; an explanation that does not refer to
measurements would be needed in the present context).  In the environment-as-a-witness paradigm, we divide $\Env$ into indivisible \emph{subenvironments}:
\begin{equation}
\Env = \Env_1 \otimes \Env_2 \otimes \Env_3 \otimes \ldots \Env_{\Nenv}.
\end{equation}

These subenvironments can be rearranged into larger \emph{fragments}.  A generic fragment consisting of $m$ subenvironments will be written as $\Envset{m}$.  The fragment containing the particular subenvironments $\{\Env_{i_1},\Env_{i_2},\ldots\Env_{i_m}\}$ is denoted $\Envset{i_1,i_2,\ldots i_m}$.

We assume that each observer captures a \emph{random} fragment of $\Env$.  This ensures their strict independence. In essence, we do not allow the observers to caucus over the partition of $\Env$, dividing it up in an advantageous way.
 
\subsubsection{How much information is practically available}

The maximum information that could be provided about $\Sys$ is its entropy, $\Hs$.  In general, no fragment can provide \emph{all} this information \footnote{The GHZ state in Eq. \ref{eqGHZ} is the exception that proves the rule.  Such states are measure-zero in Hilbert space.  Perfect C-NOT interactions are required to make them.}.  Following the reasoning in \cite{OllivierPRL04}, we demand that each fragment provide some large fraction, $1-\delta$ (where $\delta \ll 1$), of the available information about $\Sys$.  The precise magnitude of the \emph{information deficit} $\delta$ should not be important.  We denote the redundancy of ``all but $\delta$ of the available information'' by $R_\delta$.  That is, when we allow a deficit of $\delta = 0.1$, we are computing $R_{0.1}$ or $R_{10\%}$.

To compute $R_{\delta}$, we start by defining $N_{\delta}$ as \textbf{the number of disjoint fragments $\Env_i$ such that $\Isen{i} \geq (1-\delta)\Ise$}.  We might just define $R_{\delta} = N_{\delta}$, except for two caveats.
\begin{enumerate}
\item A large deficit ($\delta$) in the definition of ``sufficient'' information could lead to spurious redundancy.  Suppose there exist $N=5$ fragments that provide full information.  If $\delta = 0.5$, then we might split each fragment in half to obtain $N_{\delta}=10$ fragments that each provide ``sufficient'' information.  To compensate for this, we replace $N_{\delta}$ with $(1-\delta)N_{\delta}$.
\item Because of quantum correlations, $\Isen{i}$ can be as high as $2\Hs$.  We allow for this by \emph{assuming} that the information provided by one fragment represents strictly quantum correlations, and throwing this fragment away.  This means replacing $(1-\delta)N_{\delta}$ with $(1-\delta)N_{\delta}-1$.
\end{enumerate}
By assuming the worst case, we have obtained a \emph{lower bound} for the true redundancy:
\begin{equation}
R_\delta \geq (1-\delta) N_\delta - 1 \label{eqRedundancy}.
\end{equation}
For small $\delta$, this is fairly tight, as $N_\delta$ is clearly an upper bound.  Since our current toolset, subject to the caveats mentioned above, does not permit a more precise determination of $R_\delta$, we report the lower bound throughout.  Thus, when we report ``$R_{10\%} = 9$,'' we really mean ``$R_{10\%}$ is at least 9, and not much more.''

\subsection{Identifying qualitative redundancy} \label{secQualitativeR}

\OnePanelFigureNow[Three kinds of partial information plots (PIPs)]{figThreePIPs}{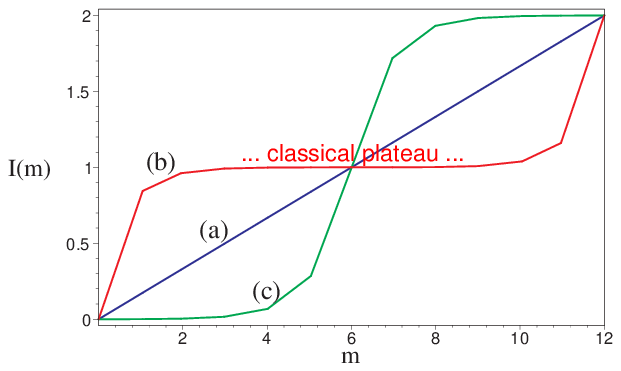}
{(Color) \textbf{Three profiles for partial information plots ($\I$ vs. $m$).}  \textbf{(a)}: the behavior of \emph{independent} environments. \textbf{(b)}: information is stored \emph{redundantly}.  \textbf{(c)}: information is \emph{encoded} in multiple environments.}

The actual \emph{amount} of redundancy is often less important than the qualitative observation that information is stored very redundantly (e.g., $R\gg1$).  Whether $R=100$ or $R=1000$, the information in question is certainly objective -- but if $R \sim 1$, then its objectivity is in doubt.  We also wish to consider more general questions: e.g., how much does $R_{\delta}$ depend on $\delta$? or \emph{why} does a state display virtually no redundancy?

For these purposes, we plot the amount of information about $\Sys$ supplied by a fragment of size $m$ ($\Isen{\{m\}}$), against $m$.  Since there are very many fragments of a given size, we average $\Isen{\{m\}}$ over a representative sample of fragments to obtain $\Ibar(m)$.  The plot of $\Ibar(m)$, which shows the partial information yielded by a partial environment, is a \emph{partial information plot} (PIP).  When the universe is in a pure state (see \cite{RBK04}, and Appendix \ref{appQMIDetails}), the PIP must be anti-symmetric around its center (see Fig. \ref{figThreePIPs}).  Together with the observation that $\Ibar(m)$ must be strictly non-decreasing (capturing \emph{more} of the environment cannot \emph{decrease} the amount of information obtained), this permits the three basic profiles shown in Figure \ref{figThreePIPs}.

Redundancy (see Fig. \ref{figThreePIPs}b) is characterized by a rapid rise of $\Ibar$ at relatively small $m$, followed by a long ``classical plateau''. In this region, all the easily available information has been obtained.  Additional environments confirm what is already known, but provide nothing new.  Only by capturing \emph{all} the environments can an observer manipulate quantum correlations.  The power to do so is indicated by the sharp rise in $\Ibar$ at $m \sim \Nenv$.

\section{Information storage in random states} \label{secUniformEnsemble}

Redundant information storage is ubiquitous in the classical world.  We might na\"{\i}vely expect that randomly chosen states of a model universe -- e.g., a $\ds$-dimensional system in contact with a bath of $\Nenv$ $\de$-dimensional systems -- would display massive redundancy.  To test this hypothesis, we compute partial information plots for random states, and average them over the uniform ensemble.  This was first done in \cite{RBK04}, for qubits.  In this work, we extend the analysis to systems and environments with arbitrary sizes.

\TwoPanelFigure[PIPs for uniform ensembles]{figUniformPIPs}{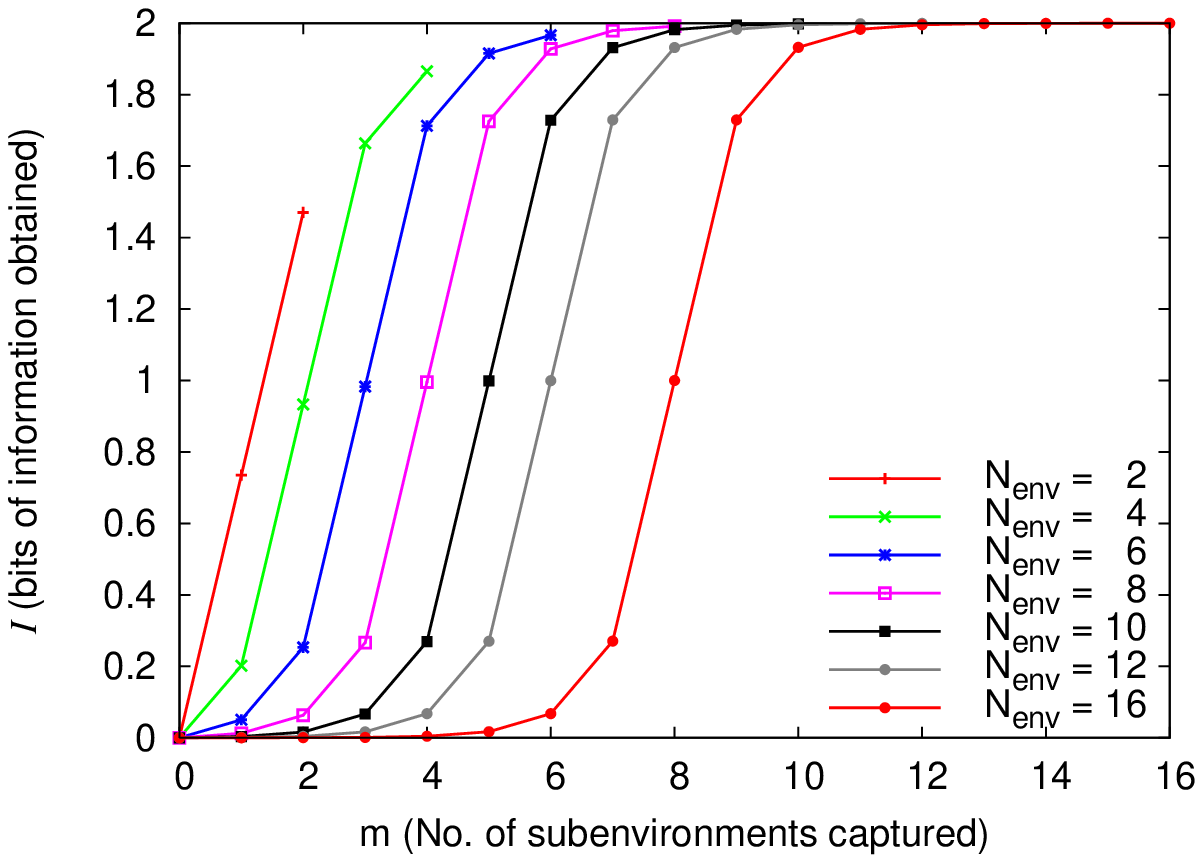}{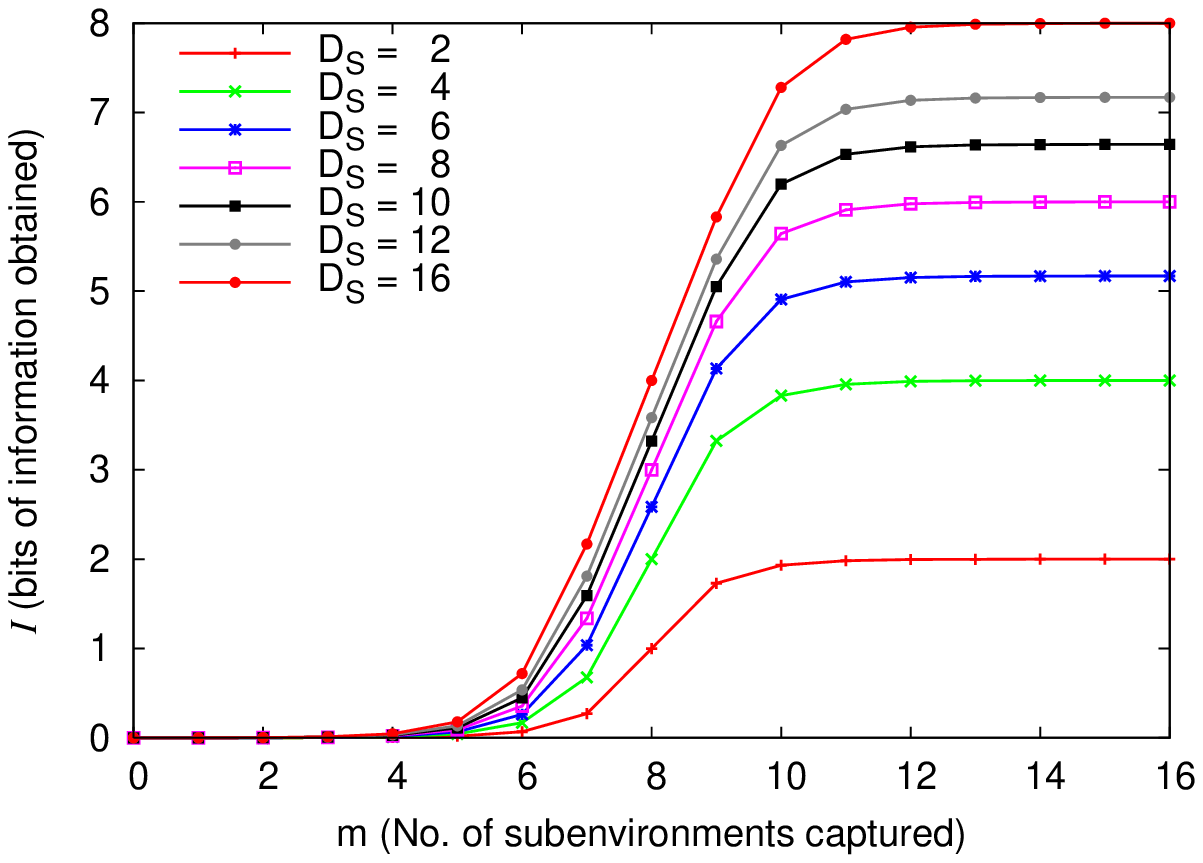}
{(Color) \textbf{Partial information plots (PIPs) for the uniform ensemble.}  We plot the average information ($\Ibar$) obtainable from a fragment ($\Envset{m}$), against the fragment's size ($m$).  $\I(m)$ is averaged over \textbf{all states in the uniform ensemble}.  \textbf{(a)}: A qubit system coupled to environments consisting of $\Nenv=2\ldots16$ qubits.  \textbf{(b)}: Systems with sizes $\ds=2\ldots16$ coupled to a 16-qubit environment. \textbf{Discussion:}  No significant information is obtained until almost half the subenvironments have been captured.  Once $m>\frac{\Nenv}{2}$, virtually all possible information (both quantum and classical) is available.  Because more than half the environment is required to obtain useful information, there is no redundant information storage in typical uniformly-distributed states.  Instead, the information is \emph{encoded} throughout the environment.}

\subsection{The uniform ensemble}

For any [finite] $D$-dimensional Hilbert space, there exists a unitarily invariant uniform distribution over states, usually referred to as \emph{Haar measure}.  We examine the behavior of typical random states by averaging PIPs over this uniform ensemble.  This average can be obtained analytically, using a formula for the average entropy of a subspace that was conjectured by Page \cite{PagePRL93}, then proved by Sen \cite{SenPRL96} and others \cite{FoongPRL94,SanchezRuizPRE95}.

Page's formula \cite{PagePRL93,SenPRL96,FoongPRL94,SanchezRuizPRE95} for the mean entropy $\overline{H}(m,n)$ of an $m$-dimensional subsystem of an $mn$-dimensional system (where $m\leq n$) is
\begin{eqnarray}
\overline{H}(m,n) &=& \sum_{k=n+1}^{mn}{\frac{1}{k}} - \frac{m-1}{2n} \\
&=& \Psi(mn) - \Psi(n+1) - \frac{m-1}{2n},
\end{eqnarray}
where the latter expression is given in terms of the \emph{digamma} $\Psi$ function.  For a $\ds$-dimensional system in contact with $\Nenv$ environments of size $\de$, the average mutual information between the system and $m$ sub-environments is
\begin{eqnarray}
\overline{\Isen{\{m\}}} =&& \overline{H}(\ds,\de^{\Nenv}) \nonumber \\
&+& \overline{H}(\de^m,\ds \de^{\Nenv-m}) \nonumber \\
&-& \overline{H}(\ds\de^m, \de^{\Nenv-m}) \label{eqHaarMI}.
\end{eqnarray}

\subsection{Partial information plots (PIPs)} \label{secUniformPIPs}

\TwoPanelFigure[Equivalent environments in the uniform ensemble]{figVariousEnvSizePIPs}{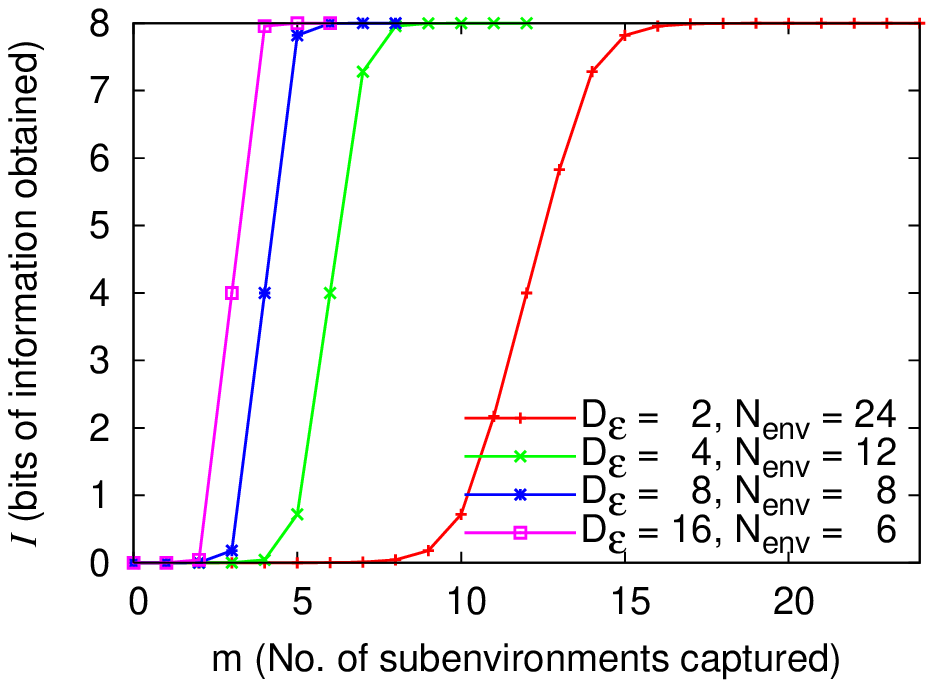}{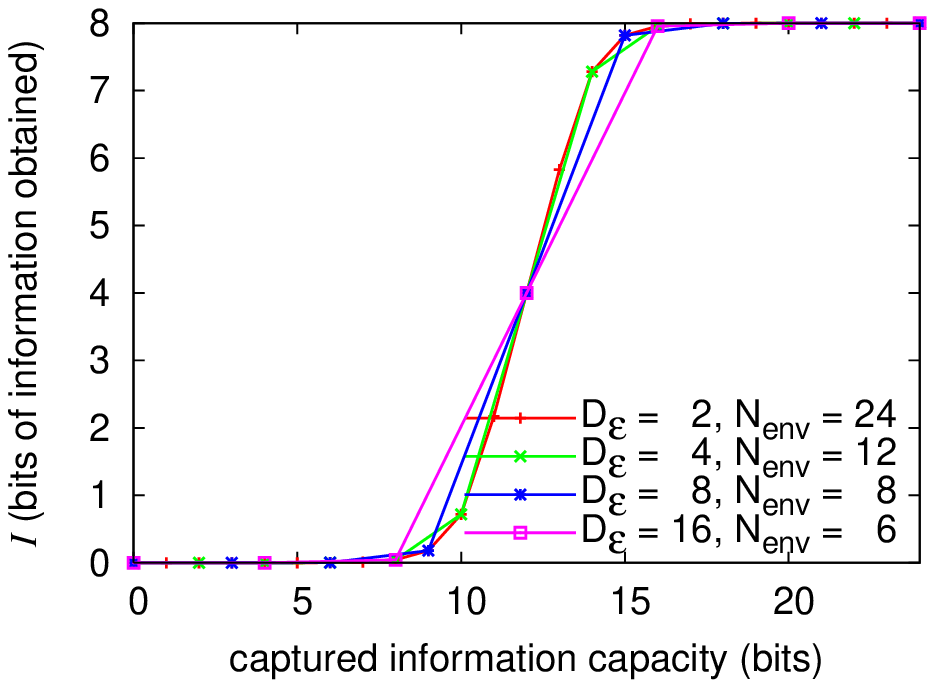}
{(Color) \textbf{Equivalent enviroments}  When the state of the universe is chosen randomly, the environment's Hilbert space dimension determines its information-recording properties.  \textbf{(a)}: PIPs for a $16$-d system coupled to several equivalent environments with $D_{\mathrm{total}} = 2^{24}$.  The subenvironments are \{2, 4, 8, 16\}-dimensional, and $\Nenv$ is scaled appropriately.  The plots are essentially identical -- only the scaling of the $m$-axis changes.  \textbf{(b)}: The same data, but with the \emph{captured fraction} of the environment plotted on the independent axis.}

\TwoPanelFigure[Scaled partial information plots (SPIPs) for the uniform ensemble]{figUniformSPIPs}{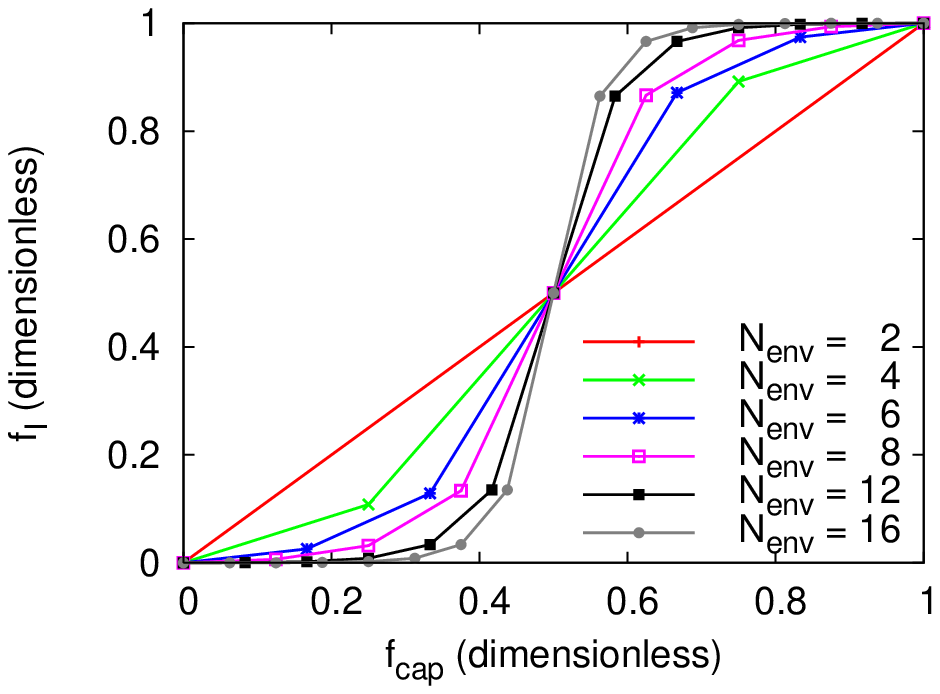}{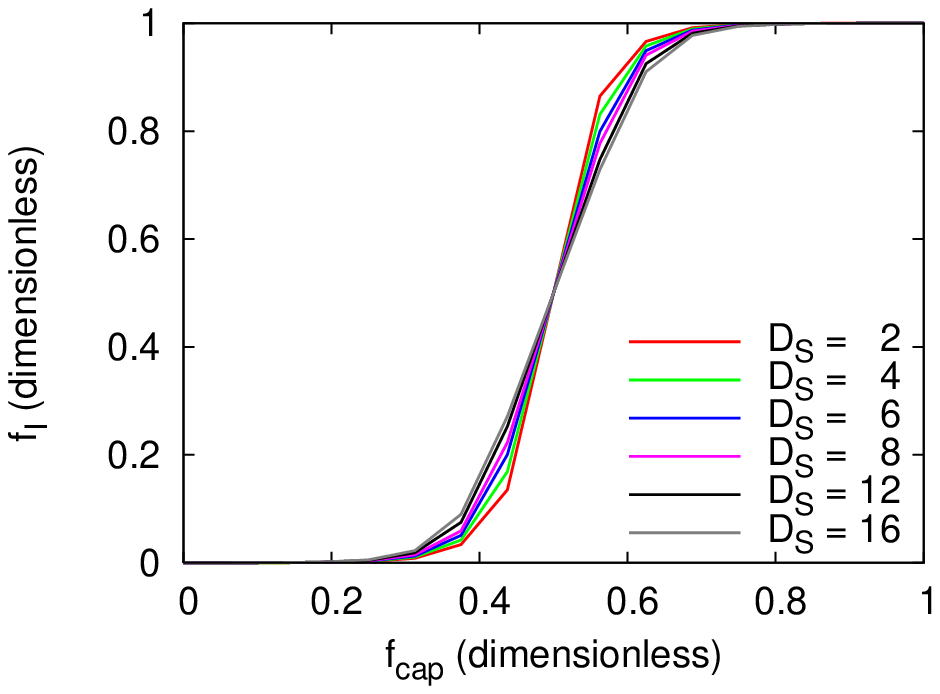}{(Color) \textbf{Scaled versions (SPIPs) of the plots in Fig. \ref{figUniformPIPs}}.  SPIPs are useful for comparing environments with different numbers of subenvironments, and for computing $R_\delta$, the redundancy for a given fraction $1-\delta$ of the total information.  To estimate redundancy, simply draw a horizontal line at $\fI = \frac{1-\delta}{2}$, and note the value of $\fcap$ where it intersects the PIP.  This provides a good estimate of $1/R_\delta$.  It is not a perfect estimate for several reasons; most importantly, the PIP and SPIP plot \emph{the average $\I$} obtained from a given-sized fragment of the environment.  This is not the same as \emph{the average fragment size} ($\overline{m}$) required to obtain $\I$, since we average the same data over different variables.  In these plots, of course, no redundancy is evident -- we are looking ahead to the next section.}

Our results (Figs. \ref{figUniformPIPs}-\ref{figUniformSPIPs}) demonstrate that \textbf{typical states from the uniform ensemble do not display redundancy}.  Figure \ref{figUniformPIPs}a illustrates typical behavior.  As an observer captures successively more subenvironments (increasing $m$), he gains virtually no information about $\Sys$.  $\Isen{\{m\}}$ remains close to zero.  When approximately $50\%$ of the subenvironments have been captured, the observer begins to gain information.  $\Ibar$ rises rapidly, through $H_s$ and onward nearly to $2H_s$.

Information about $\Sys$ is \emph{encoded} in the environment (as in Fig. \ref{figThreePIPs}c), much as a classical bit can be encoded in the parity of an ancillary bitstring.  In the classical example, however, \emph{every} bit of the ancilla must be captured to deduce the encoded bit.

This encoding, or ``anti-redundancy'', is related to quantum error correction \cite{KnillPRA97,GottesmanPRA98,NielsenBook00,ScottPRA04}.  In an encoding state, any majority subset of the $\Env_i$ has nearly-complete information.  The recorded information is unaffected by the loss of any minority subset.  States with this behavior can be used as a quantum code to protect against bit loss.  Our results show that \emph{generic} states -- i.e., states selected randomly from the whole $\Sys\Env$ Hilbert space -- form a nearly-optimal error-correction code for bit-loss errors.  Shannon noted similar behavior for classical codewords \cite{ShannonBook49}.

Figures \ref{figUniformPIPs}b and \ref{figVariousEnvSizePIPs} extend this result to larger systems.  The results are consistent; information is still encoded, and only the total amount of encoded information changes.

\subsection{Conclusions}

Our first main result is that \textbf{typical states selected randomly from the uniform ensemble display no redundant information storage}.  Instead, they display encoding or anti-redundancy.  This is not to say that \emph{all} states are ``antiredundant'', merely that redundant information storage is rare.  As $m$ declines from $\frac{\Nenv}{2}$, $\Ibar(m)$ declines exponentially.  For large $\Nenv$, states where information is \emph{not} encoded this way are vanishingly rare.  If even a small fixed fraction $\epsilon$ of states displayed the opposite ``redundant'' behavior, then $\Ibar(m)$ would have to be $O(\epsilon)$ at small $m$.  The fact that $\Ibar(m)$ is exponentially close to zero implies that the fraction of non-``encoding'' states must decline exponentially with $\Nenv$.

The obvious conclusion is that the Universe does not evolve into random states.  Our observations of ubiquitous redundancy in the real Universe are inconsistent with the random-state model.  This is interesting, but not terribly surprising.  There is no good reason to expect that the Universe's state \emph{would} be random -- we are not, for instance, in thermodynamic equilibrium.  The interactions of systems with their environments must select states that \emph{are} characterized by greater redundancy.  In the next section, we suggest and analyze such an ensemble.

\section{Decoherence and branching states} \label{secBranchingStates}

Decoherence -- the loss of information to the environment -- is a prerequisite for redundancy.  The simplest models of decoherence \cite{ZurekPRD81} are essentially identical to those for quantum measurements.  A set of pointer states for the system, $\{\ket{n}\}$, are singled out, and the environment ``measures'' which $\ket{n}$ the system is in, by evolving from some initial state ($\ket{\Env_0}$) into a \emph{conditional}  state, $\ket{\Env_n}$.  If $\rhoS$ is written out in the pointer basis, its diagonal elements ($\rho_{nn}$) remain unchanged.  Coherences between different pointer states (e.g., $\rho_{nm}$) are reduced by a \emph{decoherence factor}:
\begin{equation}
\gamma_{nm} \equiv \braket{\Env_n}{\Env_m}.
\end{equation}

We presume that (a) the subenvironments are initially unentangled, (b) each subenvironment ``measures'' the same basis of the system, and (c) the state of the universe is pure.  In this simple model, the universe is initially in a product state:
\begin{equation}
\ket{\Psi_0} = \ket{\Sys_0} \otimes \ket{\Env\idx{1}_0} \otimes \ket{\Env\idx{2}_0} \otimes \ldots \ket{\Env\idx{\Nenv}_0}.
\end{equation}
The subenvironments do not interact with each other, and the system does not evolve on its own.  Letting the system's initial state be $\ket{\Sys_0} = \sum_n{s_n\ket{n}}$, the universe evolves over time into:
\begin{equation}
\ket{\Psi_t}= \sum_n{s_n\ket{n}_\Sys\otimes \ket{\Env^{(1)}_n}\otimes \ket{\Env^{(2)}_n}\otimes \ldots\ket{\Env^{(\Nenv)}_n}}, \label{eqBranchingState}
\end{equation}
where $\ket{\Env^{(j)}_n}$ is the \emph{conditional} state into which the $j$th subenvironment evolves \emph{if} the system is in state $\ket{n}$. Different conditional states of a given subenvironment will \emph{not} generally be orthogonal to one another, except in highly simplified (e.g. C-NOT) models.

\subsection{The branching-state ensemble}

We refer to the states defined by Eq. \ref{eqBranchingState} as \emph{singly-branching states}, or simply as \emph{branching states}.  In Everett's many-worlds interpretation \cite{EverettRMP57}, a branching state's wavefunction has $\ds$ branches.  Each branch is perfectly correlated with a particular pointer state of the system.  The subenvironments are not entangled with each other, only correlated (classically) via the system.  In contrast, a typical random state from the uniform ensemble has $D_{\mathrm{universe}}$ branches, with a new branching at every subsystem.  

In dynamical models of decoherence, the universe at a given time will be described by a \emph{particular} branching state that depends on the environment's initial state, and on its dynamics.  In this \paperchapter, we sidestep the difficulties of specifying these parameters, by considering the ensemble of \emph{all} branching states.  We select the conditional $\ket{\Env_n^{(j)}}$ at random from each subenvironment's uniform ensemble.  Each pointer state of the system is correlated with a randomly chosen product state of all the environments.

The amount of available information is set by the system's initial state (i.e., the $s_n$ coefficients).  The eigenvalues of $\rhoS$ after complete decoherence, which determine its maximum entropy, are $\lambda_n = |s_n|^2$.  Since we cannot examine \emph{all} possible states, we focus on maximally ``measurable'' generalized Hadamard states:
\begin{equation}
s_n = \frac{1}{\sqrt{\ds}}\ \ \ \forall\ n.
\end{equation}
To verify that our results are generally valid, we also treat (briefly) another class of initial states.  

By examining the branching-state ensemble, we are \emph{not} conjecturing that the Universe is found exclusively in branching states.  Branching states form an interesting and physically well-motivated ensemble to explore.  We shall see that, unlike the uniform ensemble, the branching-state ensemble displays redundancy consistent with observations of the physical Universe.  Our Universe might well tend to evolve into similar states, but we are not ready to establish such a conjecture.  Characterizing the states in which the physical Universe (or a fragment thereof) \emph{is} found is a substantially more ambitious project.

\subsection{Numerical analysis of branching states} \label{secBSNumerics}

We begin our exploration of branching states by examining typical PIPs, for various systems and environments.  We average these PIPs over the branching-state ensemble, so there are only three adjustable parameters: $\ds$, $\de$, and $\Nenv$.  Our results confirm that information is stored redundantly.  Next, we examine a quantitative measure of redundancy ($R_{\delta}$), and its dependence on $\ds$, $\de$, and $\Nenv$.  Finally, we derive some analytical approximations, compare them with numerical data, and discuss the implications of our results.

\subsubsection{Partial information plots} \label{secBSPIPs}

\FourSmallPanelFigureX[PIPs for the branching ensemble]{figBSPIPs}{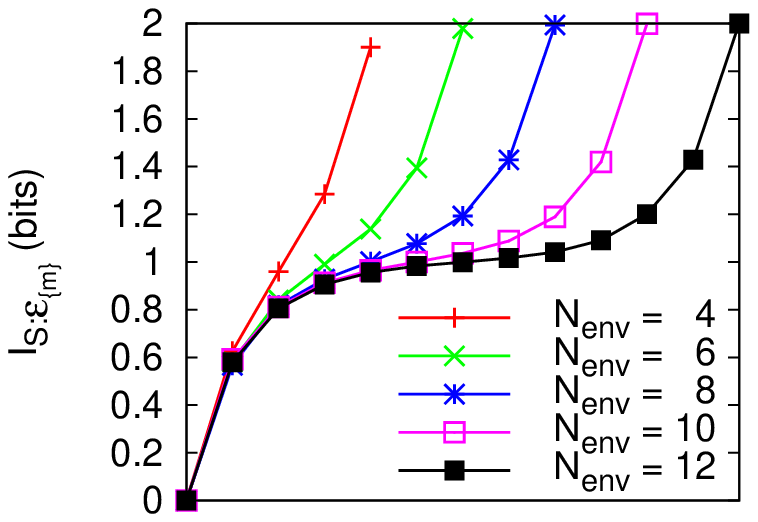}{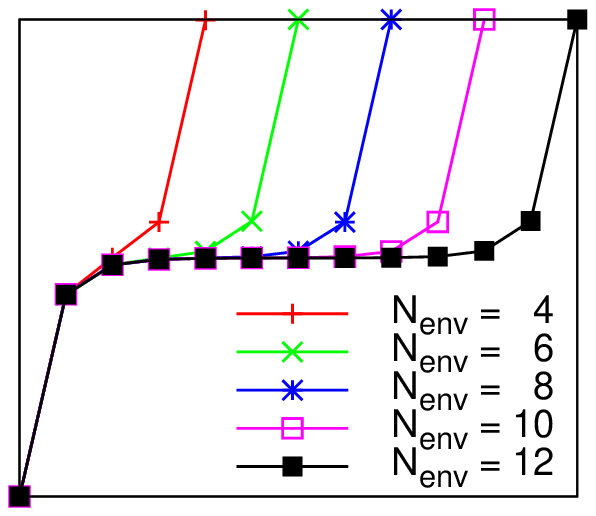}{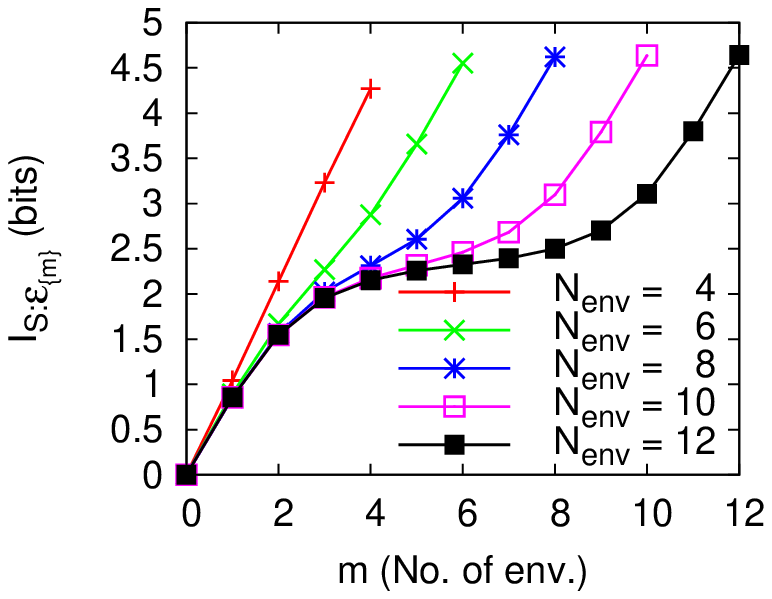}{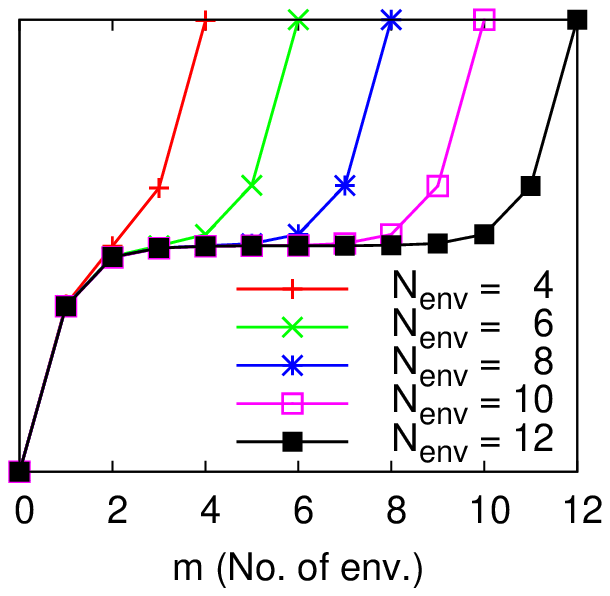}{(Color) \textbf{PIPs for ensembles of singly-branching states.}  The system is initialized in a Hadamard state, and decohered by $\Nenv$ subenvironments.  We plot the average information ($\Ibar$) available from a collection of $m$ subenvironments. \textbf{(a)}: A qubit is decohered by qubits. \textbf{(b)}: A qubit is decohered by $5$-dimensional subenvironments. \textbf{(c)}: A $5$-d system is decohered by qubits. \textbf{(d)}: A $5$-d system is decohered by $5$-d subenvironments.  \textbf{Discussion:}  As $\Nenv$ is increased from 4 to 12, a ``classical plateau'' appears.  This indicates redundant information storage.  In the regime $m\ll \Nenv$, the PIP converges to an asymptotic form.  When $\Sys$ is larger than $\Env$ (see \textbf{(c)}), the environment is barely sufficient to decohere the system, and there is no redundancy (see also Fig. \ref{figBSRedundancy}).}

Information is redundant when small fragments yield nearly-complete information -- that is, when the PIP looks like Fig. \ref{figThreePIPs}b.  PIPs for branching states (Fig. \ref{figBSPIPs}) show exactly this profile.  $\Ibar(m)$ rises rapidly from $\Ibar(0)=0$, then approaches $\Hs$ asymptotically to produce a ``classical plateau'' centered at $m = \frac{\Nenv}{2}$.

As $\Nenv$ grows, the interesting regimes at $m \sim 0$ and $m \sim \Nenv$ do not change; the classical plateau simply extends to connect them.  The initial bits of information that an observer gains about a system are extremely useful, but eventually a point of diminishing returns is reached, where further information is redundant.  The degree of redundancy should therefore scale with $\Nenv$.

\subsubsection{Non-Hadamard states for $\Sys$}

\FourSmallPanelFigureX[PIPs for inhomogeneous branching states]{figGeomPIPs}{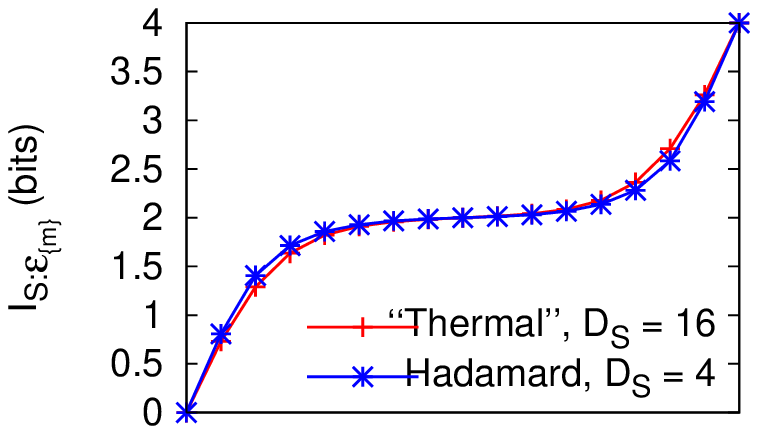}{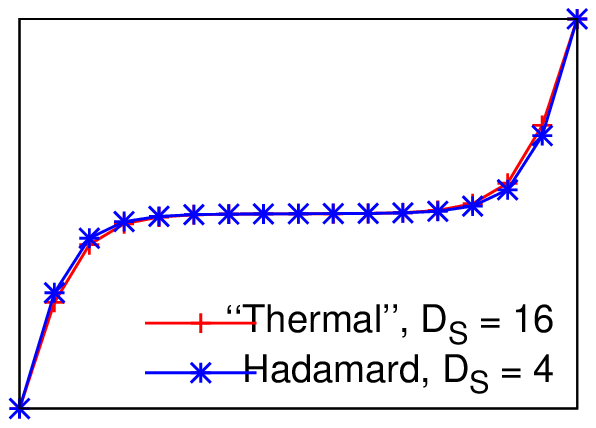}{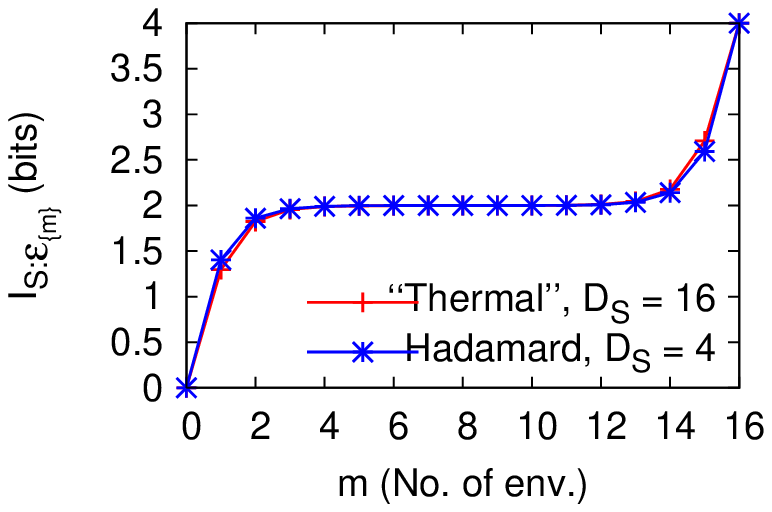}{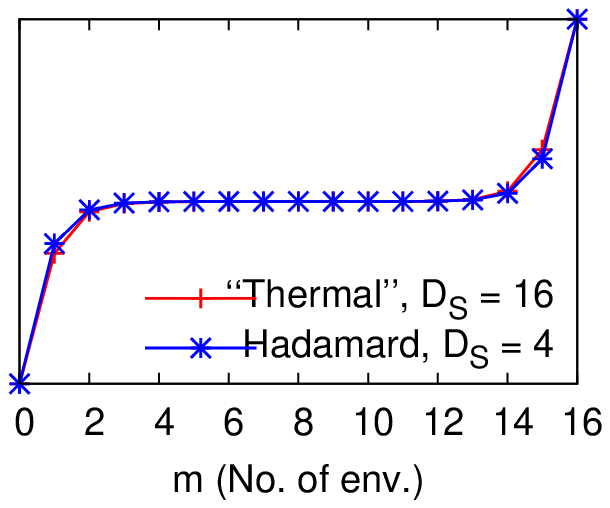}
{(Color) \textbf{PIPs for non-Hadamard states:} $\de = 2,3,4,5$ in plots \textbf{(a)}, \textbf{(b)}, \textbf{(c)}, \textbf{(d)}, respectively.  The system is $16$-dimensional, and initialized in a ``thermal'' state, where $s_n \propto \frac{1}{\sqrt{2^n}}$.  The entropy of this density matrix is $\sim2$ bits (as opposed to 4 bits for a $\ds=16$ Hadamard state).  We compare the PIPs for ``thermal'' states with $\ds=16$ to PIPs for Hadamard $\ds=4$ states, which also develop $2$ bits of entropy, varying the subenvironments' size.  These PIPs confirm that our observations apply to non-Hadamard states, and that $\Hs$ characterizes how information about the system is stored.}

Non-Hadamard states provide a different spectrum of information for $\Env$ to capture.  We consider states defined by
\begin{equation}
s_n \propto \frac{1}{\sqrt{2^n}},
\end{equation}
The post-decoherence spectrum of $\rhoS$ is non-degenerate -- in fact, it is exactly that of a thermal spin -- i.e., a particle with a Hamiltonian $\Ham = \J_z$, in equilibrium with a bath at finite temperature.  We refer to these states as \emph{``thermal''} branching states (and retain quotation marks to emphasize that our justification of this nomenclature is unphysical).

Our general approach is to assume that the system's maximum entropy determines its informational properties.  The entropy of a decohered ``thermal'' state does not increase logarithmically with $\ds$, 
but asymptotes to $\Hs=2$ bits.  This is exactly the entropy of a $\ds=4$ Hadamard state, so in the limit $\ds\rightarrow\infty$, ``thermal'' states should behave much the same as a $\ds=4$ Hadamard state.

This conjecture is confirmed in Fig. \ref{figGeomPIPs}, which compares PIPs for ``thermal'' states with $\ds=16$ to PIPs for Hadamard states with $\ds=4$.  The plots' similarity indicates that $\Hs$ is the major factor in how information about $\Sys$ is recorded.  Further numerical results use Hadamard states for specificity's sake.

\subsubsection{How PIPs scale with the composition of $\Env$}

\TwoPanelFigureNow[Scaled partial information plots (SPIPs), and the two regimes of information gain (linear and exponential)]{figBS_SPIPs}{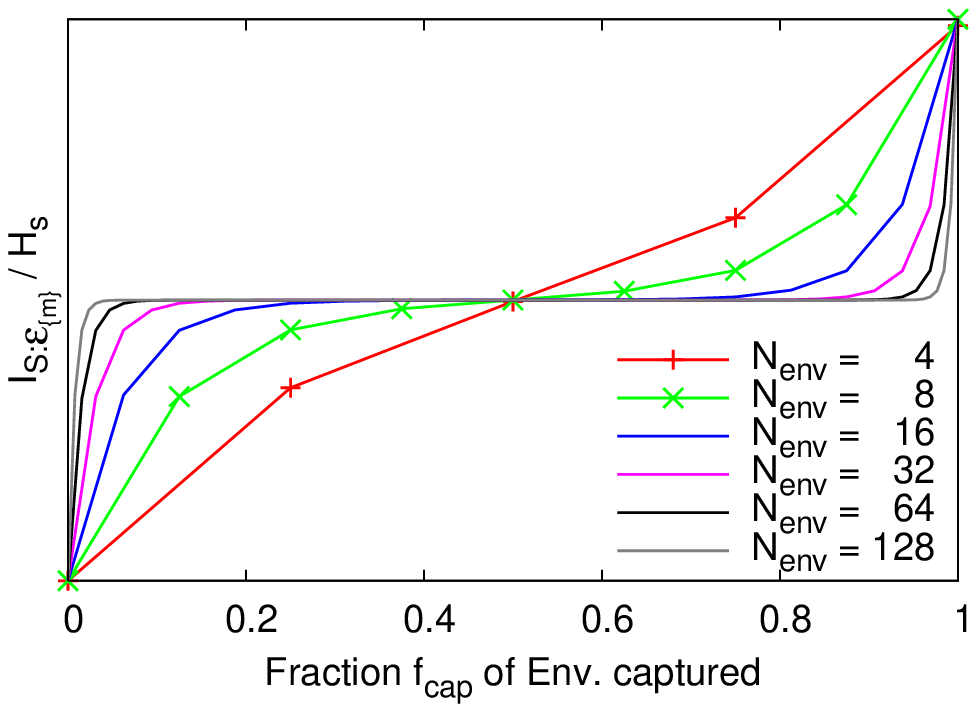} {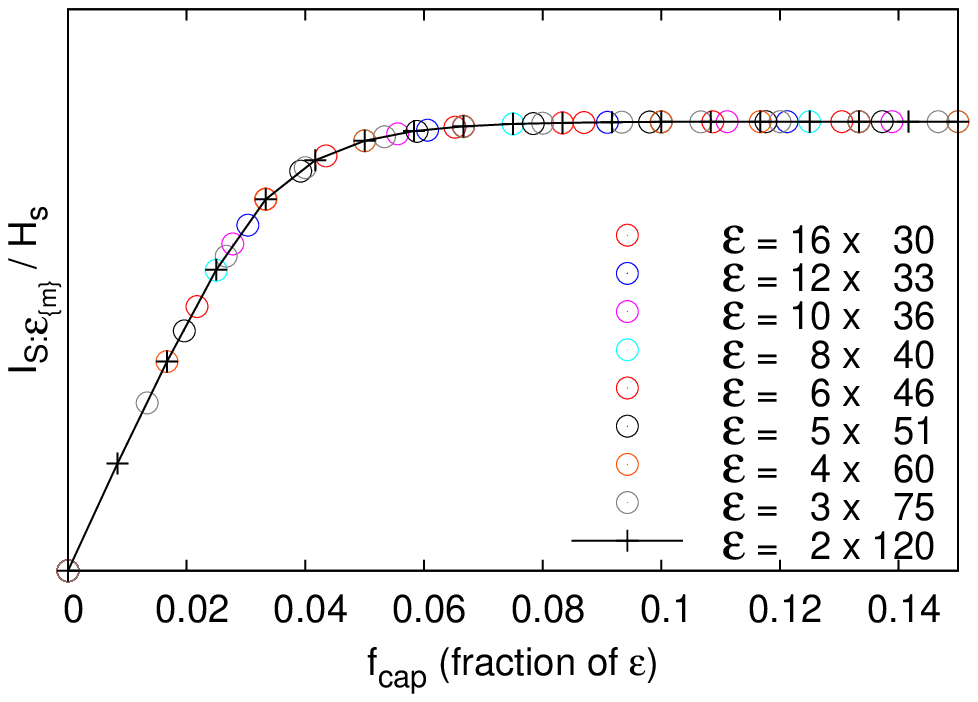}
{(Color) \textbf{Scaled partial information plots (SPIPS)} compare information storage in different environments.  \textbf{(a)}: A qutrit system coupled to $\Nenv=4\ldots128$ qutrit environments.  \textbf{(b)}: A qutrit system coupled to nine different environments with the same \emph{information capacity}.  \textbf{Discussion:} As $\Nenv$ increases, redundancy (indicated by sharp curvature) grows (plot \textbf{(a)}).  If $\Nenv$ and $\de$ are scaled so that total Hilbert space dimension ($\de^{\Nenv}$) remains constant, then the SPIP remains unchanged (plot \textbf{(b)}).  Plot \textbf{(b)} also illustrates the difference between the regime of \emph{linear} information gain (here, $f_{cap} < 0.04$) and the exponential convergence to the ``classical plateau'' thereafter.}

As the number of subenvironments in $\Env$ grows, comparing PIPs for different environments becomes difficult.  Re-parameterizing the  axes, and plotting the \emph{fraction} of $\I$ available from a \emph{fraction} of $\Env$, allows direct comparison of different universes.  Scaled PIPs (SPIPs) for environments with $\Nenv = 4\ldots128$ (Fig. \ref{figBS_SPIPs}a) show that the information about $\Sys$ becomes more redundant as $\Nenv$ grows.
 
Different environments, whose total Hilbert space dimensions are the same, act equivalently (see also Sec. \ref{secUniformPIPs}).  We have simulated a 16-dimensional system coupled to nine different, but equivalent, environments (Fig. \ref{figBS_SPIPs}b).  Although the \emph{number} and \emph{size} of the subenvironments are varied, the redundancy of the available information depends only on $\Env$'s total information capacity: $c \equiv \log\left[\mathrm{dim}\left(\mathcal{H}\right)\right]$).  Each $\Env$ in Fig. \ref{figBS_SPIPs}b has $c \simeq 120$ bits, so their SPIPs are essentially identical.

\subsubsection{Redundancy: numerical values} \label{secBSRedundancy}

\FourSmallPanelFigureX[Redundancy in branching states]{figBSRedundancy}{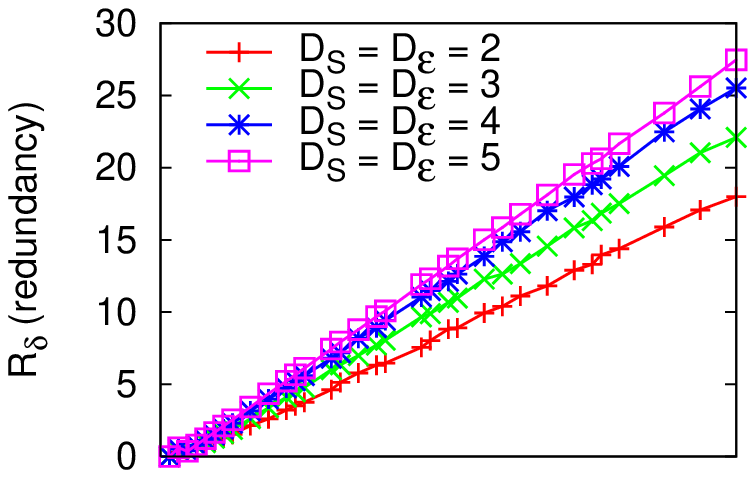}{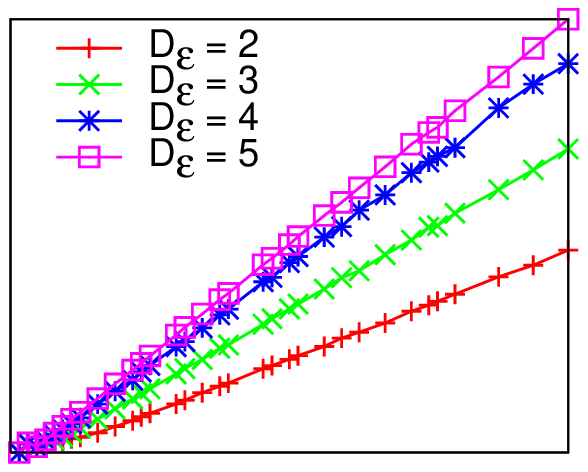}{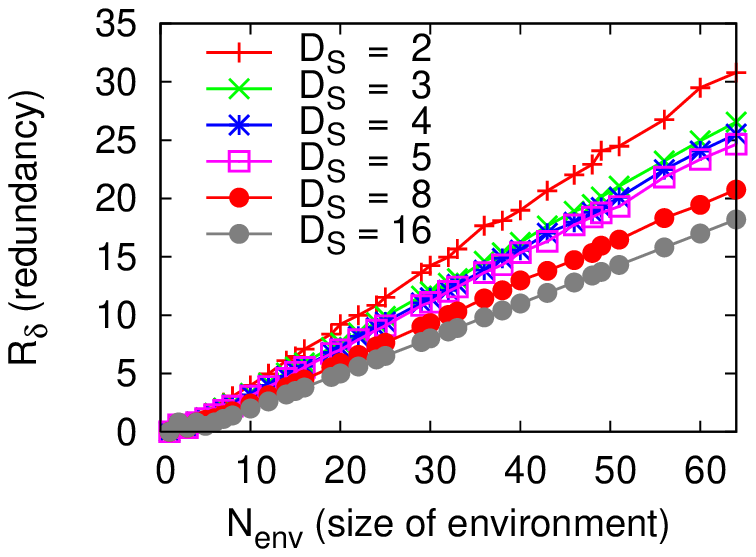} {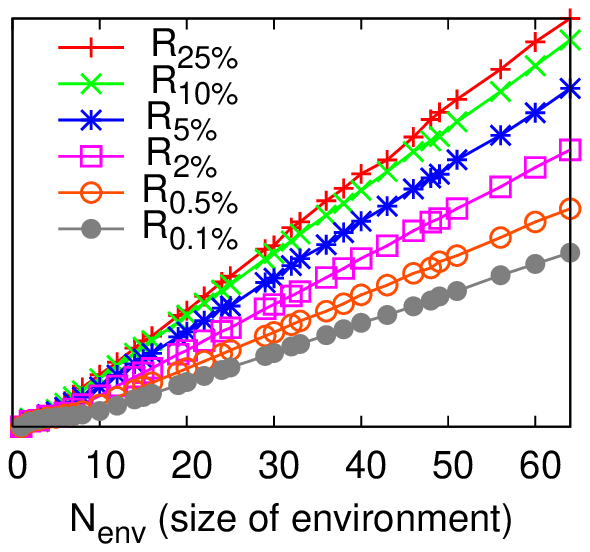}
{(Color) \textbf{Redundancy for an assortment of branching-state ensembles.}  \textbf{(a)}: $R_{10\%}$ for a $D$-dimensional system decohered by $D$-dimensional subenvironments.  \textbf{(b)}: $R_{10\%}$ for a $5$-dimensional system decohered by $\de=2\ldots5$-dimensional subenvironments.  \textbf{(c)}: $R_{10\%}$ for a $\ds=2\ldots16$-dimensional system decohered by $4$-dimensional subenvironments.  \textbf{(d)}:  $R_{\delta}$ for $\delta=0.001\ldots0.25$ and $\ds=\de=5$.
\textbf{Discussion:}  Each plot shows the ensemble-average of $R_{\delta}$, as a function of $\Nenv$.  $R_{\delta}$ increases linearly with the number of environments.  $R_{\delta}$ increases with $\de$, but decreases with $\ds$.  Larger environments store more information, which leads to greater redundancy -- but larger systems have more information to be stored.  Information is stored with slightly greater efficiency for large $\ds$ and $\de$ (plot \textbf{(a)}).  Note that if $\Sys$ is larger than $\Env$ (e.g., $\ds=16$ in plot \textbf{(c)}), there may be no redundancy.  Finally, $\delta$ affects redundancy (plot \textbf{(d)}) -- but varying $\delta$ by a full order of magnitude (from 2\% to 25\%) changes $R_{\delta}$ by less than 50\%.}

Branching states are natural generalizations of GHZ states, so we expect redundant information storage.  Figure \ref{figBSRedundancy} confirms this over a wide range of parameters.  The amount of redundancy is proportional to the size of the environment, which agrees with the classical intuition that very large environments should store many copies of information about the system.  Larger subenvironments (measured by $\de$) increase redundancy by storing more information in each subenvironment.  Conversely, larger systems have more properties to measure, which in turn require more space for information storage.  The total amount of redundancy is reduced for large $\ds$.

The other important feature of the plots in Fig. \ref{figBSRedundancy} is the relatively weak dependence of $R_{\delta}$ on the information deficit ($\delta$).  As we vary $\delta$ from $2\%$ to $25\%$ (a full order of magnitude), $R_{\delta}$ changes by less than a factor of 2.  The distinction between classical (massively redundant) and quantum (nonredundant) information is largely independent of $\delta$.

\section{Theoretical analysis of branching states} \label{secBranchingStateTheory}

The numerical analysis in the previous section offers compelling evidence that
\begin{enumerate}
\item Information is stored redundantly in branching states,
\item The amount of redundancy scales with $\Nenv$, and
\item $R_{\delta}$ is relatively insensitive to $\delta$.
\end{enumerate}
In this section, we construct theoretical models for PIPs and redundancy, which confirm these hypotheses.

\subsection{Structural properties of branching states} \label{secBSStructure}

We begin by using the structure inherent to branching states to compute a quantity of fundamental interest,
\begin{equation}
\Isen{\{m\}} = \Hs + \Hh_{\Envset{m}} - \Hh_{\Sys\Envset{m}},
\end{equation}
the mutual information between the system and a partial environment $\Envset{m}$.

We require the entropies of $\rhoS$, $\rho_{\Envset{m}}$, and $\rho_{\Sys\Envset{m}}$.  Tracing over the rest of the universe is simplified by the structure that Eq. \ref{eqBranchingState} implies.  Each relevant density matrix (regardless of its actual dimension) has only $\ds$ nonzero eigenvalues.  That is, the reduced states for $\Sys$, $\Envset{m}$, and $\Sys\Envset{m}$ are all ``virtual qudits'' with $D=\ds$.

Each $\rho$, when reduced to its $\ds$-dimensional support, is \emph{spectrally} equivalent to a partially decohered variant of the system's initial state:
\begin{equation}
\proj{\Sys_0} = \sum_{nm}{s_ns^*_m\ketbra{n}{m}}.
\end{equation}
In other words, we can obtain $\rhoS$, $\rho_{\Envset{m}}$, or $\rho_{\Sys\Envset{m}}$ by taking $\proj{\Sys_0}$ and suppressing the off-diagonal elements according to a specific rule.

To determine this rule, we define (for each subenvironment) a \emph{multiplicative decoherence factor}, $\gamma$:
\begin{equation}
\gamma\idx{k}_{ij} = \braket{{\Env^{(k)}_j}}{{\Env^{(k)}_i}},
\end{equation}
and an associated \emph{additive decoherence factor}, $d$:
\begin{equation} 
d\idx{k}_{ij} \equiv -\log\gamma^{(k)}_{ij}.
\end{equation}
Now, $\gamma\idx{k}_{ij}$ quantifies how much $\Env_k$ contributes to decohering $\ket{i}$ from $\ket{j}$.  The $\gamma$-factors from different $\Env_k$ combine multiplicatively; the $d$-factors provide a convenient additive representation.  Each relevant density matrix $\rho_X$ (for $X \in \{\Sys,\Envset{m},\Sys\Envset{m}\}$) is given by:
\begin{equation}
\braopket{i}{\rhoS}{j} = (s_is^*_j)e^{-d\idx{X}_{ij}}.
\end{equation}
The $d$-factor for each subsystem is a sum over $d$-factors for the component $\Env_k$:
\begin{eqnarray}
d^{(\Envset{m})}_{ij} &=& \sum_{k \in \Envset{m}}{d^{(k)}_{ij}} \\
d^{(\Sys)}_{ij} &=& \sum_{k \in \Env}{d^{(k)}_{ij}} \\
d^{(\Sys\Envset{m})}_{ij} &=& \sum_{k \not\in \Envset{m}}{d^{(k)}_{ij}}.
\end{eqnarray}
Thus, each $\rho$ appears to have been decohered by a different subset of $\Env$:
\begin{itemize}
\item $\rhoS$ has been decohered by \textbf{every} subenvironment,
\item $\rho_{\Sys\Envset{m}}$ has been decohered by all the subenvironments \emph{not} in $\Envset{m}$,
\item $\rho_{\Envset{m}}$ has been decohered by all the subenvironments in $\Envset{m}$.
\end{itemize}
\textbf{Note}: If the last point seems counter-intuitive, recall that for any bipartite decomposition of $\ket{\Psi}_{AB}$, the reduced $\rho_A$ and $\rho_B$ are spectrally equivalent.  Thus $\rho_{\Envset{m}}$ is equal to $\rho_{\Sys\overline{\Envset{m}}}$, where $\overline{\Envset{m}}$ contains all the environments \emph{not} in $\Envset{m}$.

Computing $\Isen{m}$ (in terms of the entropy of these three states) can be done \emph{exactly} via numerical diagonalization.  For qubit systems, it can also be done analytically (see \cite{RBK04} for extensive details).  For our model, we now derive an approximation for $\Hh(\rho)$.

\subsection{Theoretical PIPs: averaging $\I(m)$} \label{secBSApproxMI}

As a particular $\rho$ is decohered by more and more subenvironments, its off-diagonal elements decline rapidly toward zero.  We will treat the off-diagonal elements of a \emph{partially} decohered state, $\rho = \sum_{ij}{s_is^*_j\gamma_ij\ket{i}\bra{j}}$, as a perturbation around the \emph{fully} decohered state $\rho_0$, which has eigenvalues $\lambda_i = |s_i|^2$ and entropy $\Hh_0$.

\subsubsection{Average entropy of partially decohered states}

Let $\rho = \rho_0 + \Delta$, where $\Delta$ is a small off-diagonal perturbation to $\rho_0$, and expand its entropy as $\Hh(\rho) \approx \Hh(\rho_0) + O(\Delta)$.
An intuitively appealing starting point is the MacLaurin expansion of $\Hh(x) = -x\ln(x)$, which yields
\begin{equation}
\Hh(\rho_0+\Delta) \approx \Hh(\rho_0)- \Tr\left[\Delta(\Id-\ln(\rho_0))\right] - \hf\frac{\Delta^2}{\rho_0} +\frac{1}{6}\frac{\Delta^3}{\rho_0^2}\ldots 
\label{eqBadEntropyExpansion}
\end{equation}

The first order term in Eq. \ref{eqBadEntropyExpansion} vanishes, because $\Delta$ is purely off-diagonal and $\Id-\ln(\rho)$ is purely diagonal.  The leading term is thus $\frac{\Delta^2}{2\rho_0}$ -- but the matrix quotient $\frac{\Delta^{k+1}}{\rho_0^k}$ is ill-defined when $\Delta$ and $\rho_0$ do not commute.  

A more involved expansion of $\Hh(\rho)$ around $\rho=\Id$ (see Appendix \ref{appEntropyExpansion}) yields a series for $\Hh(\rho_0+\Delta)$.  It is equivalent to Eq. \ref{eqBadEntropyExpansion} for scalars, but for matrices it involves (1) expanding $\rho_0^{-k}$ in a power series, and (2) taking a totally symmetric product between $\Delta^{k+1}$ and the resulting power series.  

To leading order in $\Delta$,
\begin{equation}
\Hh(\rho) \approx \Hh(\rho_0) -\frac{\overline{|\gamma|^2}}{2}\left(h(\rho_0)-1\right),
\end{equation}
where $\overline{|\gamma|^2}$ is the average of $|\gamma_{ij}|^2$ over all $i\neq j$, and $h(\rho_0)$ is a nontrivial function,
\begin{equation}
h(\rho_0) = \sum\limits_{n,p=0}^{\infty}{\frac{\Tr\left[\rho_0(\Id-\rho_0)^p\right] \Tr\left[\rho_0(\Id-\rho_0)^{n}\right]}{n+p+1}}.\label{eqHseries} 
\end{equation}

\subsubsection{Effective Hilbert space dimension}

In general, $h(\rho_0)$ cannot be simplified further.  However, it is well approximated by the \emph{effective Hilbert space dimension} of $\rho_0$.  To see this, we consider the special case where $\rho_0$ has $D$ identical eigenvalues, $\lambda_i = \frac{1}{D}$.  When reduced to its support, $\rho_0 = \frac{\Id}{D}$.  The summation can be done explicitly:
\begin{eqnarray}
h(\rho_0) &=& \sum\limits_{n,p=0}^{\infty}{\frac{\Tr\left[\frac{\Id}{D}\left(\Id-\frac{\Id}{D}\right)^p\right] \Tr\left[\frac{\Id}{D}\left(\Id-\frac{\Id}{D}\right)^{n}\right]}{n+p+1}}\nonumber\\
&=& \sum\limits_{n,p=0}^{\infty}{\frac{\left((1-D^{-1})^p\right) \left((1-D^{-1})^{n}\right)}{n+p+1}}\nonumber\\
&=& \sum\limits_{n,p=0}^{\infty}{\frac{\left(1-D^{-1}\right)^{n+p}}{n+p+1}}\nonumber\\
&=& \sum\limits_{n+p=0}^{\infty}{\left(1-D^{-1}\right)^{n+p}}\nonumber\\
&=& D
\end{eqnarray}
Note that $D$ appeared \emph{only} based on the eigenvalue spectrum of $\rho_0$.  In the example above, the $\Hh_0 = \Hh(\rho_0) = \log(D)$.  Since the total range of $\Ibar(m)$ is proportional to $\Hh_0$, a logical generalization is
\begin{eqnarray}
h(\rho_0) &\approx& e^{\Hh_0}, \label{eqApprox_h} \\
\Hh(\rho) &\approx& \Hh(\rho_0) -\frac{\overline{|\gamma|^2}}{2}\left(e^{\Hh_0}-1\right). \label{eqApproxEntropy}
\end{eqnarray}
Numerical experimentation, and an analytic calculation in $\ds=2$, confirm that Eq. \ref{eqApprox_h} is a good approximation everywhere, in addition to being exact for (1) maximally mixed states, and (2) pure states.  

\subsubsection{Average decoherence factors}

The $\gamma_{ij}$ depend on the details of $\psi_{\Sys\Env}$.  However, when they are small enough to count as a perturbation on $\rho$, the environment's Hilbert space is very large.  The $|\gamma_{ij}|^2$ can then be treated as independent random variables, so $\overline{|\gamma^2|}$ is equal to an average over the entire branching state ensemble:
\begin{eqnarray}
\overline{|\gamma^2|} &=& \overline{\braket{\psi}{\psi'}\braket{\psi'}{\psi}} \nonumber\\
&=& \Tr\left(\overline{\proj{\psi}}\ \overline{\proj{\psi'}}\right) \nonumber\\
&=& \frac{\Tr\left(\Id\Id\right)}{\de^2} \nonumber\\
&=& \de^{-1}
\end{eqnarray}
This is the mean value of $\overline{|\gamma^2|}$ for a \emph{single} subenvironment.  For a collection of $m$ subenvironments, $m$ such $\gamma$ factors are multiplied together, so the mean value of $\overline{|\gamma^2|}$ becomes $\de^{-m}$.

\subsubsection{The result}

Putting this all together, the average entropy of a $\ds$-dimensional system decohered by $m$ $\de$-dimensional environments is
\begin{equation}
\overline{\Hh} \simeq \Hh_0 - \frac{e^{\Hh_0}-1}{2} \de^{-m}, \label{eqAverageEntropy}
\end{equation}
and the average mutual information between the system and $m$ subenvironments is
\begin{eqnarray}
\Ibar(m) &\approx& \Hh_0 - \frac{e^{\Hh_0}-1}{2}\left(\de^{-m} - \de^{-(\Nenv-m)}\right) \label{eqAverageMI} \\
&=& \Hh_0 + \left(e^{\Hh_0}-1\right)\sinh\left[\left(m-\frac{\Nenv}{2}\right)\ln(\de)\right]. \nonumber 
\end{eqnarray}

Equation \ref{eqAverageMI} is only a good approximation only near the classical plateau, where $\Ibar \simeq \Hh_0$.  Around $m=0$ and $m=\Nenv$, $\Ibar$ rises linearly, not exponentially.  Each subenvironment can provide only $\log_2\de$ bits of information, so until the information starts to become redundant, we're in a different regime (see Fig. \ref{figBS_SPIPs}b).

Once the information capacity of the captured environments ($m\log\de$) becomes greater than the amount of information in the system ($\Hh_0$), Eq. \ref{eqAverageMI} becomes valid.  It describes the slow approach to ``perfect'' information about the system, as $m$ increases.  Figure \ref{figBSApprox} compares exact (numerical) results for $\Ibar(m)$ to the approximation in Eq. \ref{eqAverageMI}. 

\SixSmallPanelFigureX[Analytical approximations to branching-state PIPs]
{figBSApprox}{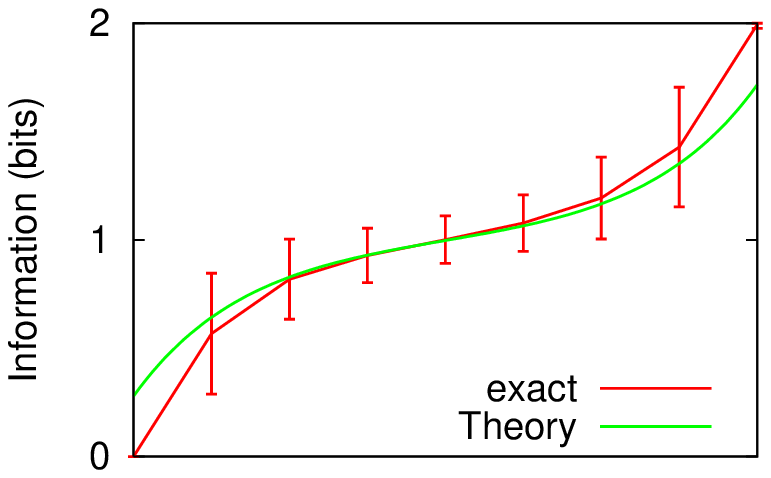}{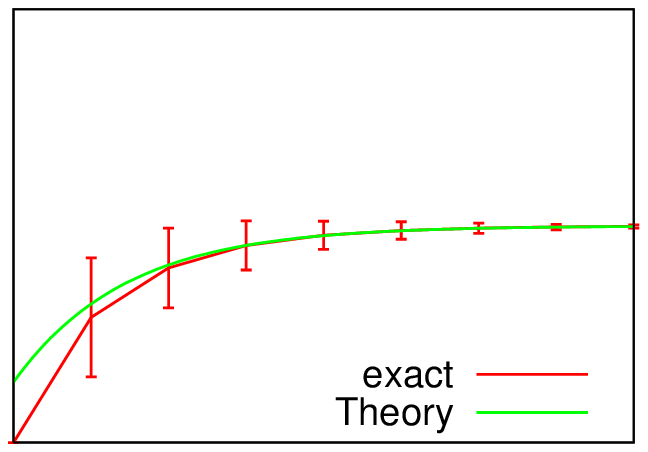}{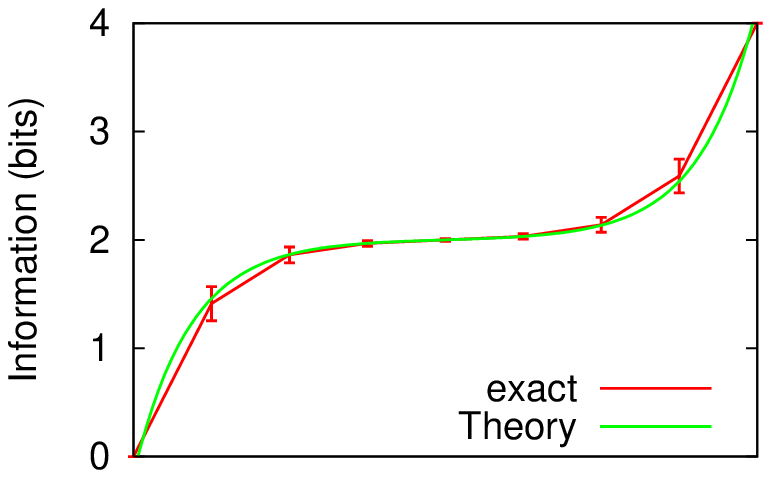}{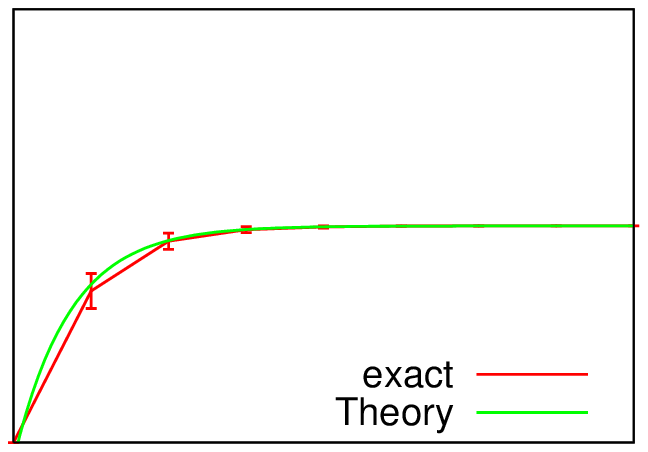}{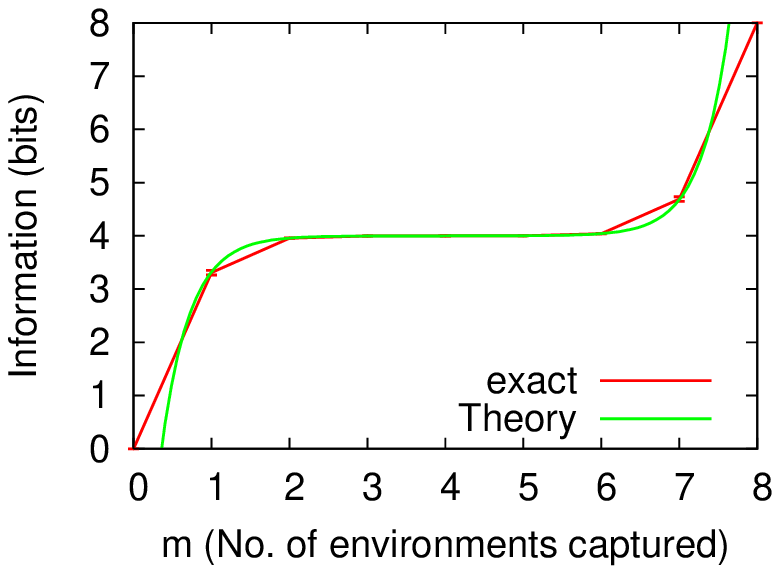}{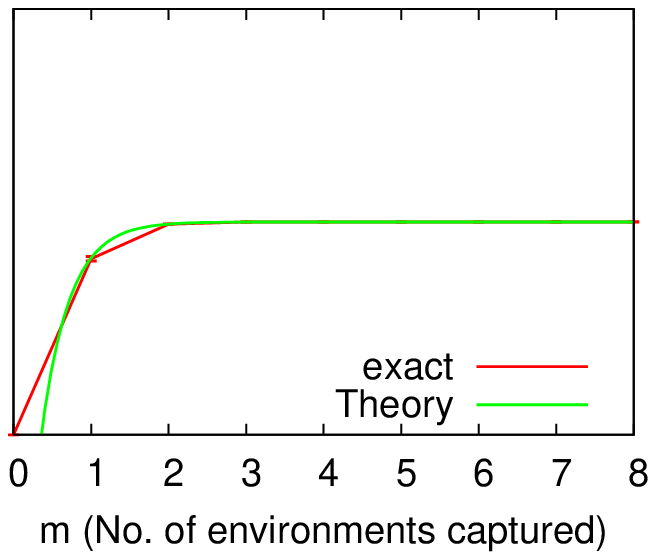}
{(Color) \textbf{Numerical PIPs vs. Theory:}  We compare the approximation derived in Sec. \ref{secBSApproxMI} with numerics.  Error bars on numerics represent typical fluctuations over the branching-state ensemble.  \textbf{(a)}: $\ds=\de=2$, $\Nenv=8$.  \textbf{(b)}: $\ds=\de=2$, $\Nenv=32$.  \textbf{(c)}: $\ds=\de=4$, $\Nenv=8$.  \textbf{(d)}: $\ds=\de=4$, $\Nenv=32$.  \textbf{(e)}: $\ds=\de=16$, $\Nenv=8$.  \textbf{(f)}: $\ds=\de=16$, $\Nenv=32$.  \textbf{Discussion:}  The approximation is virtually perfect near the classical plateau.  For small $m$, the rate of information gain is more nearly linear, and the approximation fails.  Although it works well at $m=0$ for $\ds=4$ (plots \textbf{(b)},\textbf{(e)}), it fails spectacularly near $m=0$ for large $\ds$ (plots \textbf{(c)},\textbf{(f)})}

\subsection{Theoretical redundancy: averaging $m(\I)$} \label{secBSSpecR}

Branching states develop when each subenvironment interacts independently with $\Sys$.  The data in Section \ref{secBSRedundancy} (esp. Fig. \ref{figBSRedundancy}) confirm that redundancy in branching states is proportional to $\Nenv$.  A certain number of subenvironments ($m_{\delta}$) is enough to provide sufficient information.

To capture this scaling, we define \emph{specific redundancy} as
\begin{equation}
r_{\delta} = \lim_{\Nenv\rightarrow\infty} \frac{R_{\delta}}{\Nenv} = \frac{1-\delta}{m_{\delta}} \label{eqSpecR}
\end{equation}
In this section, we use specific redundancy to examine precisely how $\ds$, $\de$, and $\delta$ affect information storage in branching states.  We derive an approximate formula for $r_{\delta}$, and compare its predictions to numerical data.

In the previous section, we computed the average information yielded by $m$ environments.  Now, we compute the average $m$ required to achieve a given $\I$.

When $\Nenv$ is large, $\Hh_{\Sys\Envset{m}} \simeq \Hs \simeq \Hh_0$, so $\Isen{\{m\}} \simeq \Hh_{\Envset{m}}$.  We take Eq. \ref{eqApproxEntropy},
\begin{equation}
\Isen{\{m\}} \approx \Hs -\hf\overline{|\gamma|^2}\left(e^{\Hs}-1\right),
\end{equation}
as a starting point.  For the fragment to provide ``sufficient'' information, $\I-\Hs$ must be less than $\delta\Hs$, which requires 
\begin{equation}
\frac{\sum_{i\neq j}{|\gamma_{ij}|^2}}{\ds(\ds-1)}\left(e^{\Hs}-1\right) \leq 2\delta\Hs.
\end{equation}
Assuming $\rho_0$ is maximally mixed (i.e., $e^{\Hh_0}=\ds$), and replacing the $\gamma_{ij}$ with independent random variables $\gamma_n$, we obtain the following condition on a ``sufficiently large'' fragment:
\begin{equation}
\left[\sum_{n=1}^{\frac{\ds(\ds-1)}{2}}{|\gamma_n|^2}\right] \leq \delta\ds\Hs \label{eqRedundancyCondition}
\end{equation}
The interaction of $\frac12\ds(\ds-1)$ independent $\gamma$-factors makes it difficult to solve Eq. \ref{eqRedundancyCondition} rigorously.  We begin instead by considering a qubit system, which has only one off-diagonal $\gamma$.

\subsubsection{Specific redundancy for qubit systems}

For a single qubit, there is only one decoherence factor: $d_{01}$, which we'll refer to simply as $d$.  Eq. \ref{eqRedundancyCondition} simplifies to:
\begin{equation}
d \geq d_{\delta} \equiv -\frac12\log\left(2\delta\Hs\right) \label{eqQubitRedundancyCondition}
\end{equation}
The increase in $d$ with $m$ can be approximated as a biased random walk, where each step has a mean length ($\overline{d}$) and a variance ($\Delta d$).  After $m$ environments are added to the fragment, $d$ obeys a normal distribution ($p_m(d)$), whose mean and variance are $m\overline{d}$ and $\sqrt{m}\Delta d$, respectively.  We postpone the calculation of $\overline{d}$ and $\Delta d$ for the moment.  

Let $\psuff(m)$ be the probability that a fragment consisting of $m$ subenvironments provides sufficient information (i.e., satisfies equation \ref{eqQubitRedundancyCondition}).  Then
\begin{equation}
\psuff(m) = \int_{d_{\delta}}^{\infty}{p_m(d)d\!d},
\end{equation}
and the probability that $m$ environments are \emph{required} is
\begin{eqnarray}
\preq(m) &=& \psuff(m) - \psuff(m-1) \\
&=& \int_{m-1}^{m}{\pdiff{}{n}\psuff(n)d\!n},
\end{eqnarray}
and the expected fragment size ($\overline{m}$) is
\begin{eqnarray}
\overline{m} &=& \sum_{m=0}^{\infty}{m\ \preq(m)} \nonumber\\
&=& \sum_{m=0}^{\infty}{m\int_{m-1}^{m}{\pdiff{}{n}\psuff(n)d\!n}} \nonumber\\
&\simeq& \int_{0}^{\infty}{\left(m+\frac12\right)\pdiff{}{m}\psuff(m)d\!m} \nonumber\\
&=& \frac12 + \int_{0}^{\infty}{m\pdiff{}{m}\psuff(m)d\!m} \nonumber\\
&=& \frac12 + \int_{0}^{\infty}{\left(1-\psuff(m)\right)d\!m} \nonumber\\
&=& \frac12 + \int_{0}^{\infty}{d\!m\int_{-\infty}^{d_{\delta}}{p_m(d)d\!d}}. \label{eqMeanMIntegral}
\end{eqnarray}
We interchange the order of integration, substitute the appropriate normal distribution for $p_m(d)$, and end up with
\begin{equation}
\overline{m} = \frac{d_{\delta}}{\overline{d}} + \frac{\Delta^2}{2\overline{d}^2}+\frac12. \label{eqMeanMPrelim}
\end{equation}

\subsubsection{Specific redundancy for general $\ds$}

Whereas Eq. \ref{eqQubitRedundancyCondition} (for qubits) has one $\overline{|\gamma|^2}$ term, Eq. \ref{eqRedundancyCondition} involves a sum of $\frac12\ds(\ds+1)$ such terms.  Deriving an analyzing a probability distribution for this sum is very difficult, so we take a simpler route.  We replace the sum over terms with a single term, $\frac12\ds(\ds+1)\cdot\gamma^2$, where $\gamma^2$ represents all the off-diagonal terms.  The new condition for sufficient information is:
\begin{eqnarray}
\frac{\ds(\ds-1)}{2}{\gamma^2} &\leq& \delta\ds\Hs \nonumber\\
\gamma^2 &\leq& \frac{2\delta\Hs}{\ds-1} \nonumber\\
d &\geq& d_{\delta} \equiv -\frac12\log\left(\frac{2\delta\Hs}{\ds-1}\right).
\end{eqnarray}
$\ds$ has been incorporated into a redefinition of $d_{\delta}$.  Equation \ref{eqMeanMPrelim} is still valid for qubits, but it generalizes to
\begin{equation}
\overline{m} = \frac{\log(\ds-1)-\log\left(2\delta\Hs\right)}{2\overline{d}} + \frac{\Delta d^2}{2\overline{d}^2} +\frac12. \label{eqMeanM}
\end{equation}
We combine this expression with Eq. \ref{eqSpecR} to obtain a general estimate for specific redundancy:
\begin{equation}
r_{\delta} = \frac{2\overline{d}^2(1-\delta)}{{\Delta}^2 + \overline{d}^2 + \overline{d}\left(\log(\ds-1)-\log(2\delta\Hs)\right)} \label{eqApproxSpecR}
\end{equation}

\subsubsection{Dependence of mean decoherence factor ($d$) on $\de$}

The computation of $\overline{d}$ and $\Delta d$ in terms of $\de$ is somewhat tedious.  Details can be found in Appendix \ref{appDFactors}, where we calculate:
\begin{eqnarray}
\overline{d} &=& \frac12\left(\Psi(\de)+\EMConst\right), \label{eqMeanD}\\
\Delta d^2 &=& \frac{\pi^2}{24} - \frac{\Psi_1(\de)}{4}, \label{eqVarianceD}
\end{eqnarray}
in terms of the digamma ($\Psi(n)$) and trigamma ($\Psi_1(n)$ functions \cite{MathworldDigamma,MathworldTrigamma}, and the Euler-Mascheroni constant $\EMConst = 0.577\ldots$.  These functions may not be familiar to all readers, so we present the first few values in Table \ref{tabDValues}.

\begin{table}[hdt!]
\begin{center}
\renewcommand{\arraystretch}{1.5}
\begin{tabular}{|l||c|c|c|c|c|c|}\hline
$\de$ & 2 & 3 & 4 & 5 & 6 & 8\\
\hline
$\overline{d}$ & $\frac12$ & $\frac34$ & $\frac{11}{12}$ & $\frac{25}{24}$ & $\frac{137}{120}$ & $\frac{363}{280}$\\
[0.5ex]\hline
$\Delta d$ & $\frac12$ & $\frac{\sqrt{5}}{4}$ & $\frac{7}{12}$ & $\frac{\sqrt{205}}{24}$ & $\frac{\sqrt{5269}}{120}$ & $\frac{\sqrt{266681}}{840}$\\
[0.5ex]\hline
\end{tabular}
\end{center}
\caption[Additive decoherence factors for small environments]{\label{tabDValues}The table shows the first few values of $\overline{d}$ and $\Delta d$, for environments of size $\de \in [2,3,4,5,6,8]$.  See Appendix \ref{appDFactors} for details on the calculation.}
\end{table}

For larger $\de$, we can safely approximate Eqs. \ref{eqMeanD}-\ref{eqVarianceD} as:
\begin{eqnarray}
\overline{d} &\simeq& \frac12\left(\log(\de)+\EMConst\right) \\
\Delta d &\simeq& \frac{\pi}{\sqrt{24}}.
\end{eqnarray}

\subsubsection{How good is the estimate?}

\FourPanelFigure[Specific redundancy in branching-state ensembles]{figSpecR}{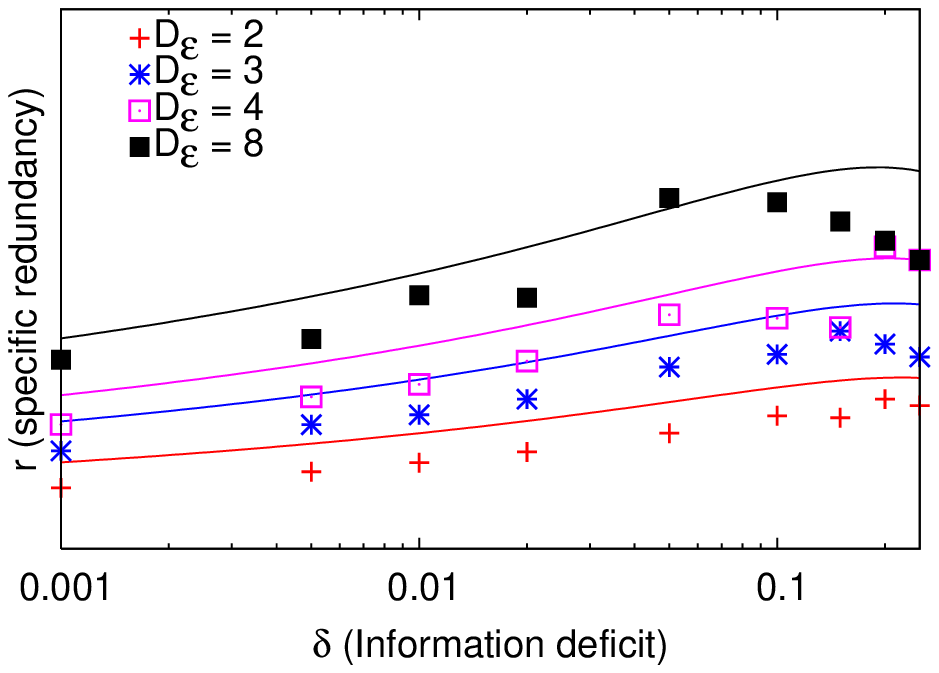}{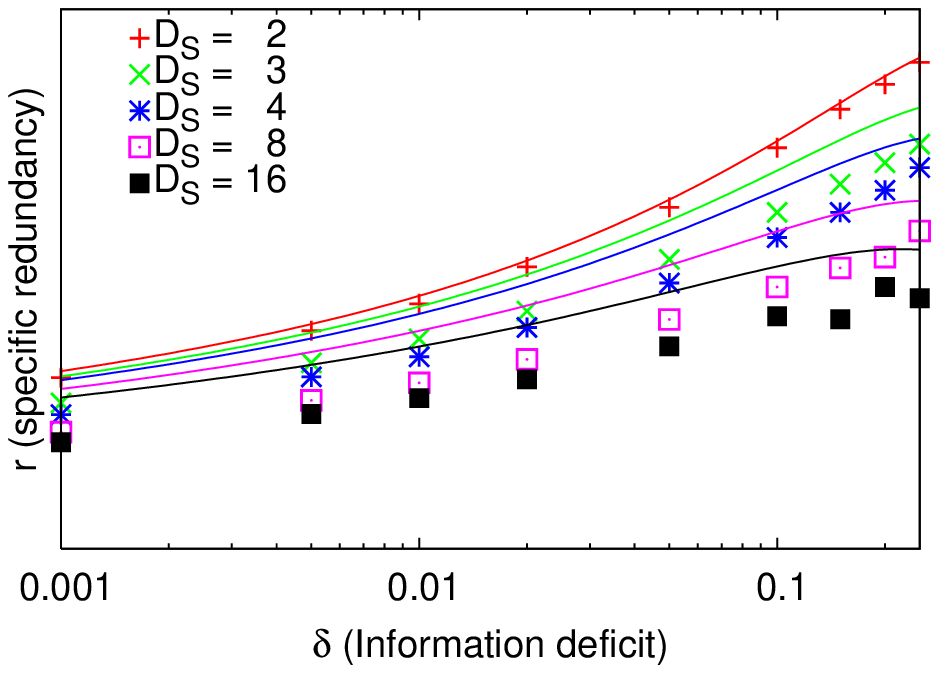}{SpecR_NxN_0.99a.\grext}{SpecR_NxN_0.99b.\grext}
{(Color) \textbf{Specific redundancy} ($r_{\delta} \equiv R_{\delta}/\Nenv$): numerical data (symbols) compared with theory (Eq. \ref{eqApproxSpecR}, solid lines).  \textbf{(a)}: $r_{\delta}$ vs. $\delta$, for a $16$-d system coupled to $2,3,4,8$-dimensional subenvironments.  \textbf{(b)}: $r_{\delta}$ vs. $\delta$, for $2,3,4,8,16$-d systems coupled to qubit subenvironments.  \textbf{(c)}: $r_{1\%}$ vs. $\de$.  \textbf{(d)}: $r_{1\%}$ vs. $\ds$.  \textbf{Discussion:}  Theory predicts the overall behavior of redundancy well.  It is nearly perfect for $\ds=2$, but overestimates $r$ for larger systems.    As $\delta$ increases, $r_{\delta}$ saturates and even \emph{declines} because of the $(1-\delta)$ prefactor in Eq. \ref{eqRedundancy}.  When $\delta$ is large, the theory breaks down (see \textbf{(a)}), because a single subenvironment can provide sufficient information.}

In Figure \ref{figSpecR}, we compare numerical results to the approximation of Eq. \ref{eqApproxSpecR}.  The analytical estimate is very good for qubit systems, but loses some fidelity for larger $\ds$.  A more sophisticated treatment of the multiple $\gamma_{ij}$ terms -- each representing an independent observable which the environment must record -- would eliminate this error.

To get an intuitive feel for the dependence of $r_{\delta}$ on its parameters, we consider the regime of large systems, large environments, and small deficit -- i.e., $H_0 \gg 1$, $\overline{d} \sim \frac12 \log(\de)$, $\Delta d \sim \frac{\pi^2}{24}$, and $\delta \ll 1$.  In this regime, we can ruthlessly simplify Eq. \ref{eqApproxSpecR} to obtain a simple prediction:
\begin{equation}
r_{\delta} \approx \frac{\log(\de)}{\log(\ds)-\log(\delta)}. \label{eqThumbnailSpecR}
\end{equation}
The plots in Fig. \ref{figEfficiency} show the ratio between numerical $r_\delta$ data and the simple predictions of Eq.\ref{eqThumbnailSpecR}.  They confirm that Eq.\ref{eqThumbnailSpecR} is a good rule of thumb.

\FourSmallPanelFigure[``Efficiency'': specific redundancy rescaled by information capacity] {figEfficiency}{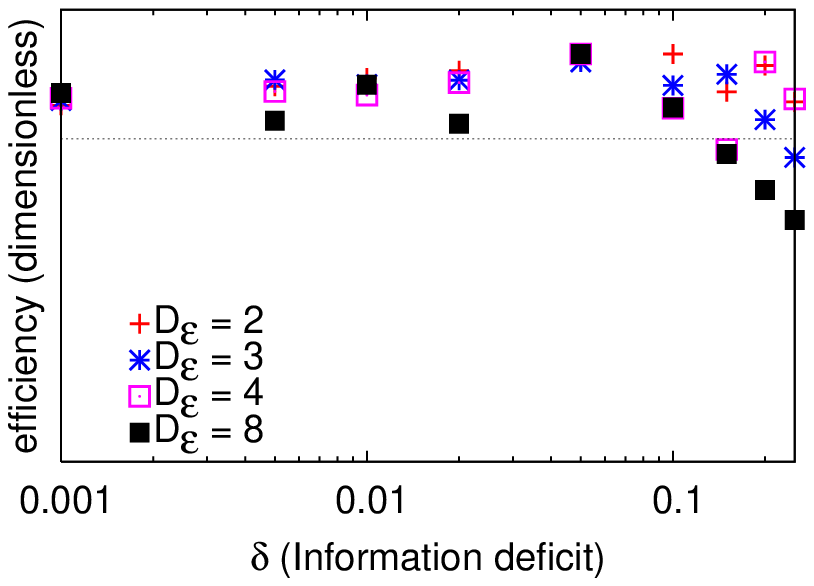}{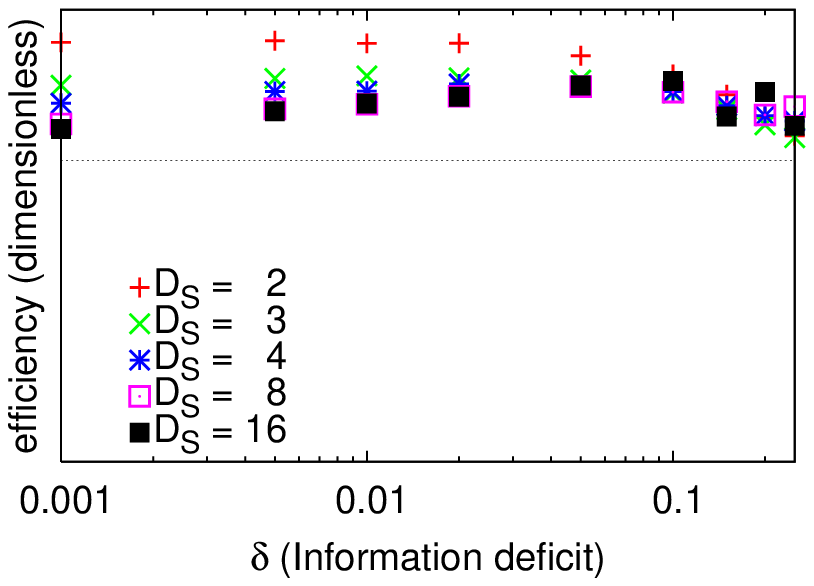}{Efficiency_NxN_0.99a.\grext}{Efficiency_NxN_0.99b.\grext}
{(Color) \textbf{``Efficiency'': specific redundancy rescaled by information capacity.}  Equation \ref{eqThumbnailSpecR} provides a simple approximation for redundancy, based on the relative information capacity of the system (with a correction for $\delta$) and its environment.  We reproduce the data of Fig. \ref{figSpecR}, but use Eq. \ref{eqApproxSpecR} to rescale specific redundancy.  \textbf{Discussion:}  Efficiency is consistently near to 1:  when the universe is in a random branching state, information about $\Sys$ is efficiently recorded in $\Env$.  Equation \ref{eqApproxSpecR} is accurate for large $\ds$ and $\de$ (and small $\delta$).  When the system or the subenvironments are small, Eq. \ref{eqApproxSpecR} underestimates information storage efficiency.}

Eq. \ref{eqThumbnailSpecR} can be interpreted as a capsule summary of how redundancy scales in the ``random-state'' model of decoherence.
\begin{enumerate}
\item Redundancy is proportional to $\Nenv$, the number of independent subenvironments.  \textbf{More environments produce more redundancy.}
\item Redundancy is proportional to $\bar{d}$, the \emph{mean decoherence factor} of a single subenvironment, which grows as $\log\de$.  \textbf{Larger environments produce more redundancy, in proportion to their information capacity.}
\item Redundancy is (roughly) inversely proportional to $\Hs$, the total information available about the system.  \textbf{Larger systems require more space in the environment.}
\item The deficit ($\delta$) appears as a logarithmic addition to $\Hs$.  Reducing the amount of ``ignorable'' information is equivalent to making the system bigger.  \textbf{Redundancy depends only weakly (logarithmically) on the deficit, $\delta$.}
\end{enumerate}

\section{Conclusions and discussion} \label{secStaticsConclusions}

\OnePanelFigure[Quantum Darwinism in action]{figInfoDivision}{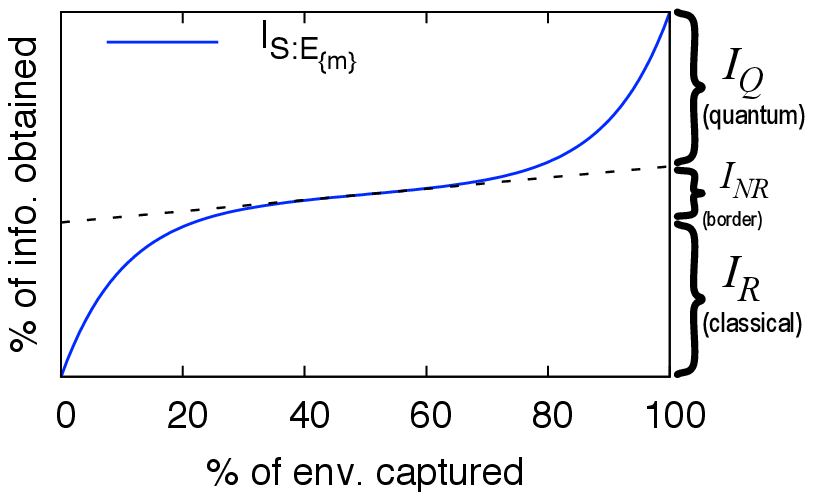}
{(Color) \emph{Quantum Darwinism} selects certain observable properties of the system and propagates information about them throughout the environment.  The preferred observable[s] become redundant \textbf{at the expense of incompatible observables}.  As shown here, PIPs illustrate the results of Quantum Darwinism.  Information about $\Sys$ becomes divided into three parts: redundant information ($\Ir$), quantum information ($\Iq$), and non-redundant information ($\Inr$).  Redundant information is objective, and therefore classical.  It can be obtained with relative ease.  Quantum information represents the non-preferred observables, marginalized by Quantum Darwinism, which can only be measured by capturing all of $\Env$.  Non-redundant information (determined by the slope of $\Ibar(m)$ at $m=\frac{\Nenv}{2}$) represents the ambiguous borderline, undifferentiated as yet into classical and quantum fractions.  When $\Inr$ is small, the central region of the PIP becomes flat.  This ``classical plateau'' indicates that an observer can obtain full information without capturing the entire environment.}

`There is no information without representation': information has to be stored somewhere.  To retrieve it, we must measure the systems where it is stored.  To understand the properties of information, we look at the properties of this retrieval process.  We have focused on the question: \textbf{How easily can information about a system be retrieved from its environment?}

The answer is strongly dependent on how the system became correlated with its environment.  Random interactions between $\Sys$ and all of $\Env$ leave no useful correlations -- to learn about $\Sys$ we must measure most of $\Env$.  However, when localized parts of $\Env$ interact independently with $\Sys$, an observer can learn about $\Sys$ by measuring a small fragment of $\Env$.  Furthermore, the information that he learns is objective -- another independent observer will arrive at the same conclusions.

This redundant imprinting of selected observables on the environment is \emph{quantum Darwinism}.  It leads to objective reality in a quantum Universe.  Typical PIPs for branching states (see Fig. \ref{figInfoDivision}) illustrate how different sorts of information are selected or deprecated.  The information in $\Env$ about $\Sys$ divides naturally into three parts.
\begin{equation}
\Ise = \Ir + \Inr + \Iq.
\end{equation}
The \emph{redundant information} ($\Ir$) is classical -- it can be obtained easily, by many independent observers. Its selective proliferation is the essence of quantum Darwinism.  Ollivier et. al. showed, in \cite{OllivierPRL04}, that $\Ir$ is not only easy to obtain, but difficult to ignore.  An observer who succeeds in extracting $\Ir$, and continues to probe, finds a ``classical plateau''.  Measurements on additional subenvironments increase his knowledge of $\Sys$ only slightly -- mostly, they only confirm what he already knows.  Only a \emph{perfect} and \emph{global}  measurement of \emph{everything} can reveal more than the redundant information.

Purely \emph{quantum information} ($\Iq$) represents observables that are incompatible with the pointer observable.  This is the information that quantum Darwinism selects \emph{against}.  It is (a) encoded amongst the environments, much as a classical bit can be encoded in the parity of many ancilla bits; (b) accessible only through a global measurement on \emph{all} of $\Env$; and (c) easily destroyed when $\Env$ decoheres.

Finally, \emph{non-redundant information} ($\Inr$) represents a grey area -- the border between the classical and quantum domains. It exists only when the classical plateau in $\Ibar(m)$ has a nonzero slope.  This is why we allow for a deficit ($\delta$) when computing redundancy. 


Information storage in randomly selected \emph{arbitrary} states of the model universe is dramatically different from information storage in randomly selected \emph{singly-branching} states.  The contrast between these two cases emphasizes the importance of the environment's structure.  Overly simple thermodynamic arguments (e.g., maximum entropy in absence of gravity) indicate that the physical Universe should evolve into states that are uniformly distributed.  Our results, however, show that objects which display the redundancy characteristic of \emph{our} Universe must have \emph{structured} correlations with their environments.  

Decoherence theory emphasizes the role of the environment in the quantum-to-classical transition, but only as a reservoir where unwanted quantum superpositions and correlations can be hidden, out of sight.  Even this view -- which now seems somewhat narrow -- has produced important advances in our understanding over the past quarter century.  Examples include einselection, the special role of pointer states, and the view of classicality as an emergent phenomenon.  Nevertheless, it is clear from our discussion above and from related recent work \cite{OllivierPRL04,Ollivier04}, that ``tracing out $\Env$'' obscures crucial aspects of the environment's role.

\textbf{The environment is a witness} -- a communication channel through which observers acquire the vast majority (if not all) of their information about the Universe.  Surprisingly, this realization has taken more than 75 years since the formulation of quantum mechanics in its present form.  It goes against a strong classical tradition of looking for solutions of fundamental problems in {\it isolated} settings.  This tradition is incompatible with the role of \emph{states} in quantum theory.

Quantum states, unlike classical states, do not define what ``exists objectively''. They are too malleable -- too easily perturbed and redefined by measurements.  Moreover, in quantum mechanics, what is \emph{known} about a system's state is inextricably intertwined with what it \emph{is}.  Classical states, in contrast, have existence independently of the knowledge of them.  To put it tersely (and in the spirit of complementarity), quantum states play both ontic (describing what is) and epistemic (describing what is known to be) roles\footnote{See \cite{ZurekRMP03} and especially \cite{Zurek04} for further discussion of quantum states' {\it epiontic} nature.}.  Thus, for many purposes, it makes no sense to talk about a state of a completely isolated quantum system.

Our Universe is `quantum to the core' (see e.g. Ref. \cite{Schlosshauer05b} for an up-to-date review of the experimental evidence), so the only place to look for objective classicality is within the quantum theory itself. Decoherence has certainly supplied part of the answer: Only \emph{some} of the states in an open system's Hilbert space are stable.  Those that are not stable, cannot ``exist objectively''.  Even these einselected pointer states, however, are vulnerable to perturbation by an observer who measures directly.  Yet, objectivity implies that many different (and initially ignorant) observers can independently find out the state. 

The environment-as-a-witness point of view solves this problem by recognizing that we gain essentially all of our information indirectly, from the environmental degrees of freedom (with the possible exception of specific laboratory experiments). As the environment is the ``channel'', and as only a part of it can be intercepted, the obvious question is: \textbf{How is information is deposited in ${\cal E}$? and what kind of information?}

Quantum Darwinism, which we have begun to analyse here and elsewhere \cite{ZurekRMP03,OllivierPRL04,Ollivier04,ZurekADP00}, aims to supply the answer. Our basic conclusion is that the redundancy evident in our Universe is not a generic property of randomly selected states in large multipartite (system plus multi-component environment) Hilbert spaces.  However, when states in that Hilbert space are created by the interactions usually invoked in discussions of environment-induced superselection, redundancy appears.  Thus, objectivity can arise through the dynamics of decoherence.  In that sense, decoherence is the mechanism that delivers quantum Darwinism -- a more complete view of classicality's emergence.

While we have already witnessed the birth of this new point of view, it is still far from mature. In particular, our conclusion about redundancy and the typical structure of entanglement was reached without analyzing dynamics {\it per se}.  We have laid the foundation for a full-fledged study of quantum Darwinism by analysing kinematic properties of states, and postponed the study
of evolution in specific models to forthcoming publications \cite{RBK05c,RBK05e}.  Moreover, by employing von Neumann entropy, we have focused on the amount of information (rather than on what this information is about). Differences between various definitions of mutual information exist (see ``discord", Ref. \cite{OllivierPRL02}), and are symptomatic of the ``quantumness'' of the underlying correlations. Less ``quantum'' definitions of mutual information, involving conditional information, {\it de facto} presume a measurement.  They have also been used \cite{ZurekRMP03,OllivierPRL04,Ollivier04}), along with other tools (\cite{DalvitPRL01, Dalvit05}), to show that the familar pointer observables are the ``fittest'' in the (quantum) Darwinian sense. Studying the dynamics of quantum Darwinism, and the connections with various definitions of information, are the obvious next steps.

\acknowledgments
We thank Harold Ollivier and David Poulin for vigorous discussions.
This research was supported in part by NSA and ARDA.

\appendix

\section{Properties of QMI: the Symmetry Theorem} \label{appQMIDetails}

The \emph{symmetry theorem} for QMI is important for understanding the shape of PIPs (partial information plots).  It says, in essence, that the amount of information that can be gained from the \emph{first} few environments to be captured, is mirrored by the amount of information that can be gained from the \emph{last} few environments.  Thus, when capturing a small fraction of $\Env$ yields much information, an equivalent amount of information \emph{cannot} be gained without capturing the last outstanding bits of $\Env$.

\begin{theorem}[Mutual Information Symmetry Theorem]\label{thmMISymmetry}
Let the universe be in a pure state $\ket{\psi}_{\Sys\Env}$, and let the environment $\Env$ be partitioned into two chunks $\Env_A$ and $\Env_B$.  Then the total mutual information between the system and its environment is equal to the sum of the mutual informations between $\Sys$ and $\Env_A$ and between $\Sys$ and $\Env_B$: that is, $\Ise = \Isen{A} + \Isen{B}$.\end{theorem}
\textbf{Proof.}  We simply expand each mutual information as $\I_{x:y} = H_x + H_y - H_{xy}$, and use the fact that if a bipartite system $x\otimes y$ has a pure state $\ket{\psi}_{xy}$, then the entropies of the parts are equal; $H_x = H_y$.
\begin{eqnarray*}
\Isen{A} + \Isen{B} &=& H_S + H_A - H_{SA} + H_S + H_B - H_{SB}\\
&=& H_S + H_A - H_B + H_S + H_B - H_A\\
&=& H_S + H_S\\
&=& H_S + H_{AB} - 0\\
&=& \Ise
\end{eqnarray*}

\begin{corollary}\label{thmTwoEnvironmentLimit} Under no circumstances can two sub-environments \emph{both} have $\I > \Hs$ information about the system.\end{corollary}
If the universe is in a pure state, then the Symmetry Theorem states that any bipartite division of the environment will yield two chunks, at least one of which has $\I \leq \Hs$.  Additionally, we note that a chunk has at least as much $\I$ about the system as any of its sub-chunks (that is, decreasing the size of a chunk cannot increase its $\I$).  If we could find two chunks $A$ and $B$ with $\I > \Hs$, then by subsuming the remainder of $\Env$ into $A$ we would have a bipartite division into $A'$ and $B$, each of which has $\I > \Hs$ -- but this contradicts the Symmetry Theorem.

The proof for a mixed state of the universe follows from the ``Church of the Larger Hilbert Space'' argument.  We purify $\rhoSE$ by enlarging the environment from $\Env$ to $\Env'$, and follow the same steps to show that $\Env'$ cannot have two subenvironments with $\I > \Hs$.  Since $\Env$ is a subset of $\Env'$, it too cannot have two such subenvironments.

\begin{corollary}\label{thmPIPSymmetry} For a pure state $\ket{\psi}_{\Sys\Env}$ of the universe, the partial information plot (PIP) must be antisymmetric around the point $(m=\frac{N}{2},\I = \Hs)$. \end{corollary}

This follows straightforwardly from the Symmetry Theorem.  For each chunk $\Envset{m}$ of the environment that contains $m$ individual environments, there exists a complementary chunk $\Envset{N-m}$, containing the complement of $\Envset{m}$, with $N-m$ individual environments.  The Symmetry Theorem implies that $\Isen{\{m\}} + \Isen{\{N-m\}} = \Ise = 2\Hs$.  By averaging this equation over all possible chunks $\Envset{m}$, we obtain an equation for the PIP: $\Ibar(m) + \Ibar(N-m) = 2\Hs$.  This equation is equivalent to the stated Corollary.

\section{Perfect states}

The primary intuition that we obtain from the $\Ibar(m)$ plots is that most states are ``encoding'' states, but an important sub-ensemble of states are ``redundant'' states.  We are naturally led to ask whether ``perfect'' examples of each type of state exist -- that is, a state that encodes information more redundantly than any other state, or a state that hides the encoded information better than any other state.

The answer is somewhat surprising: whereas perfectly redundant states exist for any $N$ and any $\ds, \de$, perfect coding states apparently exist only for certain $N$ (at least for $\ds=\de=2$).  The perfectly redundant states are easy to understand; they are the generalized GHZ (and GHZ-like) states of the form:
\begin{equation}
\ket{\Psi_{\Sys\Env}} = \alpha\ket{0}_{\Sys}\Tensor_i{\ket{0}_{\Env_{i}}}
			+ \beta\ket{1}_{\Sys}\Tensor_i{\ket{1}_{\Env_{i}}}, \label{eqGHZLikeStates}
\end{equation}
with the obvious generalizations to higher $\ds,\de$.  Of course, it's necessary that $\de \geq \ds$.

A true GHZ state is invariant under interchange of any two subsystems; however, since mutual information is invariant under local unitaries, we only require that the states $\ket{0}_{\Env_{i}}$ and $\ket{1}_{\Env_{i}}$ be orthogonal.  Clearly, such states exist for all $N$.  Any sub-environment with $0<m<N$ has exactly $H(\Sys)$ information, but only by capturing the entire environment ($m=N$) can we obtain the full $\I = 2H(\Sys)$.  Thus, the information is stored with $N$-fold redundancy.

A perfect coding state, on the other hand, would be one where $\Ibar(m)=0$ for any $m<N/2$, and $\Ibar(m)=\I_{\Sys\Env}$ for $m>N/2$.  An equivalent condition, for qubit universes, is the existence of two orthogonal states of $N$ qubits, each of which is maximally entangled under all possible bipartite divisions.  If such pairs of states exist, then the system states $\ket{0}$ and $\ket{1}$ can be correlated with them to produce the perfect coding state.  It is known (as detailed in \cite{ScottPRA04}) that such states only exist for $N=2,3,5,6$, and possibly for $N=7$ (for $N=6$, only a single state exists\cite{CalderbankIEEE98}).  Thus, while for large $N$ almost every state is an excellent coding state, perfect examples seem not to exist except for $N=2,3,5,(7?)$!  We are not aware of any results for non-qubit systems.

\section{Entropy of a near-diagonal density matrix} \label{appEntropyExpansion}

Suppose that the pure state $\oper{\pi} = \proj{\psi}$, whose components in the pointer basis are
\begin{equation}
\braket{i}{\psi} = s_i,
\end{equation}
is subjected to decoherence.  The off-diagonal elements are reduced according to
\begin{equation}
\pi_{i,j} \longrightarrow \sigma_{i,j} = \gamma_{i,j}\pi_{i,j},
\end{equation}
where $\gamma_{i,i}=1$ for all $i$.  The limiting point of the process, where $\gamma_{i,j} = 0$ for all $i\neq j$, is $\rho$:
\begin{equation}
\rho_{i,j} = \delta_{ij}|s_i|^2. \label{eqDiagonal}
\end{equation}

As the $\gamma_{i,j}$ approach zero, $\sigma$ converges to $\rho$.  The partially decohered $\sigma$ can be written as
\begin{equation}
\sigma = \rho + \Delta,
\end{equation}
where $\Delta$ is strictly off-diagonal.  $\Delta$ is defined by
\begin{equation}
\Delta_{i,j} = \left(1-\delta_{ij}\right) \gamma_{i,j}s_is_j^*. \label{eqOffdiagonal}
\end{equation}
As $\sigma$ approaches $\rho$, its entropy approaches the entropy of $\rho$.  Our goal here is to write $\Hh(\sigma)$ as a power series (in $\Delta$) around $\Hh(\rho)$.

The entropy of $\sigma$ is
\begin{equation}
\Hh(\sigma) = -\Tr(\sigma\ln\sigma) = \Tr\left(\Hsquig(\sigma)\right)
\end{equation}
where 
\begin{equation}
\Hsquig(\sigma) \equiv -\sigma\ln\sigma. \label{eqHsquig}
\end{equation}
The difference between $\Hh(\sigma)$ and $\Hh(\rho)$ is
\begin{equation}
\dH = \Tr(\dHsquig) = \Tr\left(\Hsquig(\rho+\Delta) - \Hsquig(\rho)\right).
\end{equation}
We will seek a power series for $\dHsquig$.  Keeping in mind that its \emph{trace} is the relevant quantity, we will discard traceless terms.

\subsection{A na\"{\i}ve approach to expanding $\Hh(\rho+\Delta)$}

It's tempting to begin by expanding Eq. \ref{eqHsquig} around $\sigma=\rho$.  Using the MacLaurin series for $-\sigma\ln\sigma$ gives
\begin{eqnarray}
\Hsquig &=& -\Delta\left(\Id+\ln\rho\right) - \sum\limits_{n=0}^{\infty}{\frac{(-1)^n}{(n+1)(n+2)}\frac{\Delta^{n+2}}{\rho^{n+1}}} \\
&\approx& - \frac{\Delta^2}{2\rho} + \frac{\Delta^3}{6\rho^2} \ldots. \label{eqBadSeries}
\end{eqnarray}
We discarded the first term because it is traceless.
Unfortunately, matrix quotients are not well-defined.  $\frac{\Delta}{\rho}$ could mean either $\Delta\rho^{-1}$ or $\rho^{-1}\Delta$ -- and, in fact, both are nonsymmetric and therefore incorrect.  Other symmetric orderings, such as $\rho^{-\hf}\Delta\rho^{-\hf}$, also give incorrect results.  The expansion in Eq. \ref{eqBadSeries} is an inappropriate generalization of a \emph{scalar} expansion, and is ill-defined.  We will take a different approach which (a) gives the correct result, and (b) defines the correct representation of matrix quotients.

\subsection{The correct approach}

Instead of expanding $\Hsquig(\sigma)$ around $\sigma=\rho$, we expand \emph{both} $\Hsquig(\sigma)$ and $\Hsquig(\rho)$ around the identity.
\begin{eqnarray}
\dHsquig &=& \Hsquig(\rho+\Delta) - \Hsquig(\rho) \nonumber \\
&=& \Hsquig(\Id - (\Id-\rho-\Delta)) - \Hsquig(\Id - (\Id-\rho)). \nonumber
\end{eqnarray}
The expansion around $\Id$ is always well-defined, because $\Id$ and its inverse commute with everything:
\begin{equation}
\Hsquig(\Id-x) = x-\sum\limits_{n=0}^{\infty}{\frac{x^{n+2}}{(n+1)(n+2)}}.
\end{equation}
Using this expansion in $\dHsquig$ yields
\begin{equation}
\dHsquig = - \Delta + \sum\limits_{n=0}^{\infty}{\frac{\left[(\Id-\rho)^{n+2} - (\Id-\rho-\Delta)^{n+2}\right]}{(n+1)(n+2)}}.
\end{equation}
We once again discard $\Delta$ because it is traceless, leaving only the sum.  The two matrix powers within the sum can be rewritten using the identity
\begin{equation}
(\Id+x)^n = \sum\limits_{j=0}^{n}{\binom{n}{j}x^n},
\end{equation}
which yields
\begin{equation}
\dHsquig = -\sum\limits_{n=0}^{\infty}{\sum\limits_{j=0}^{n+2}{(-1)^j\binom{n+2}{j}\left((\rho+\Delta)^j-\rho^j\right)}}.
\end{equation}

In order to simplify this, we must introduce a new notation.  Consider $(x+y)^p$, where $x$ and $y$ may be either scalars or matrices.  For scalar $x$ and $y$, 
\begin{equation}
(x+y)^p = \sum\limits_{k=0}^{p}{\binom{p}{k}a^k b^{p-k}},
\end{equation}
whereas for matrices, $\binom{p}{k} x^k y^{p-k}$ is replaced by a sum over $\binom{p}{k}$ orderings of $k$ $x$'s and $p-k$ $y$'s.  We define the notation $x^k \spr y^{p-k}$ to describe this sum: e.g.,
\begin{equation}
x^2 \spr y^2 = \frac{x^2y^2 + xyxy + xy^2x + yx^2y + yxyx + y^2x}{6},
\end{equation}
but when $x$ and $y$ are scalars
\begin{equation}
x^2 \spr y^2 = x^2 y^2.
\end{equation}
Using this definition of a \emph{totally symmetric product},
\begin{equation}
(\rho + \Delta)^j = \sum\limits_{k=0}^{j}{\binom{j}{k} \Delta^k \spr \rho^{j-k}},
\end{equation}
and the entropy difference operator $\dHsquig$ is

\widetextstart
\begin{eqnarray}
\dHsquig &=& -\sum\limits_{n=0}^{\infty}{\sum\limits_{j=0}^{n+2}{\sum\limits_{k=0}^{j-1}
	{\frac{(-1)^j}{(n+1)(n+2)}\binom{n+2}{j}\binom{j}{k+1}\Delta^{k+1}\spr\rho^{j-k-1}}}}
	\label{eqdHsquigRaw} \\
&=& -\sum\limits_{k=1}^{\infty}{\sum\limits_{n=0}^{\infty}{\sum\limits_{j=0}^{n}
{\frac{(-1)^{j+k+1}}{(n+k)(n+k+1)}\binom{n+k+1}{j+k+1}\binom{j+k+1}{k+1}\Delta^{k+1}\spr\rho^{j}}}} \\
&& - {\sum\limits_{n=0}^{\infty}{\sum\limits_{j=0}^{n+1}{\frac{(-1)^j}{(n+1)(n+2)}(j+1)\binom{n+2}{j+1}\Delta \spr \rho^j}}}
\end{eqnarray}
The $k=0$ term can be discarded because $\Tr(\Delta\spr\rho^j)=\Tr(\Delta\rho^j)=0$.  We then perform the sum over $j$ to obtain
\begin{equation}
\dHsquig = -\sum\limits_{k=1}^{\infty}{\sum\limits_{n=0}^{\infty}
{\frac{(-1)^k}{(n+k)(n+k+1)}\binom{n+k+1}{k+1} \Delta^{k+1}\spr(\Id-\rho)^n}}.
\end{equation}
\widetextend

Expanding the binomial coefficients and simplifying leads to the following result:
\begin{equation}
\dHsquig = \sum\limits_{k=1}^{\infty}{\frac{(-1)^k}{k(k+1)} \Delta^{k+1} \spr \sum\limits_{n=0}^{\infty}{\binom{k+n-1}{n}(\Id-\rho)^n}}.\label{eqdHsquig}
\end{equation}
We have come full circle.  The sum over $n$ in Eq. \ref{eqdHsquig} is just the MacLaurin expansion for $\rho^{-k}$ around $\rho=\Id$.  Equation \ref{eqdHsquig} can thus be written \emph{symbolically} as
\begin{equation}
\dHsquig = \sum\limits_{k=1}^{\infty}{\frac{(-1)^k}{k(k+1)} \Delta^{k+1} \spr \left[\rho^{-k}\right]},
\end{equation}
if the symmetric product $\Delta^{k+1} \spr \rho^{-k}$ is interpreted as ``take the symmetric product of $\Delta^{k+1}$ with the power series representing $\rho^{-k}$.''

Essentially, what we have derived is the ``correct'' interpretation of the matrix quotient $\frac{\Delta^{k+1}}{\rho^k}$.  This result is interesting in its own right, but for now we are interested only in the leading order (i.e., $\Delta^2$) term.  Truncating the series at $k=1$, we obtain the following simple result:
\begin{equation}
\dH \approx -hf\sum\limits_{n=0}^{\infty}{\Tr\left[\Delta^2 \spr (\Id-\rho)^n\right]} + O\left(\Delta^3\right). \label{eqdH}
\end{equation}
This is the simplest possible \emph{general} form for $\dH$.  In order to perform the traces, we need to take advantage of the form of the symmetric product.

From the definition of the symmetric product, we can write out explicit expressions for $\Delta^k \spr M^n$, for particular small values of $k$.
\begin{equation}
\Delta \spr M^n = \frac{1}{n+1}\sum\limits_{p=0}^n{M^p\Delta M^{n-p}}
\end{equation}
\begin{equation}
\Delta^2 \spr M^n = \frac{2}{(n+1)(n+2)} \sum\limits_{p=0}^n{ \sum\limits_{q=0}^{n-p}{M^q\Delta M^p\Delta M^{n-p-q}}}
\end{equation}
The second case (for $\Delta^2$) is the useful one.  We need the trace of the symmetric product, which can be simplified using the cyclic property of trace,
\begin{equation}
\Tr\left[\Delta^2 \spr M^n\right] = \frac{1}{n+1}\sum\limits_{p=0}^n{\Tr\left[\Delta M^p \Delta M^{n-p}\right]}.
\end{equation}
Together with Eq. \ref{eqdH}, this formula yields an \emph{explicit} expression for $\dH$:
\begin{equation}
\dH \approx -\hf\sum\limits_{n=0}^{\infty}\frac{1}{n+1}\sum\limits_{p=0}^n{\Tr\left[\Delta (\Id-\rho)^p\Delta (\Id-\rho)^{n-p}\right]} 
\label{eqdHexplicit}
\end{equation}

\widetextstart
We now insert specific forms for $\rho$ and $\Delta$, from Eqs. \ref{eqDiagonal} and \ref{eqOffdiagonal}:
\begin{eqnarray}
\Tr\left[\Delta M^p \Delta M^{n-p}\right] &=& \sum\limits_{i,j,k,l=0}^{\ds-1}{\Delta_{ij}(\Id-\rho)^p_{jk}\Delta_{kl}(1-\rho)^{n-p}_{li}} \\
&=& \sum\limits_{i,j,k,l=0}^{\ds-1}{s_i s_j^* s_k s_l^* \gamma_{ij}\gamma_{kl}\delta_{jk}\delta_{il}
	\left(1-|s_j|^2\right)^p\left(1-|s_i|^2\right)^{n-p}} \\
&=& \sum\limits_{i,j\neq i}{|s_i|^2\left(1-|s_i|^2\right)^{n-p}|s_j|^2\left(1-|s_j|^2\right)^p|\gamma_{ij}|^2}.
\end{eqnarray}
Since the goal is to average over an ensemble of states, we replace $|\gamma_{ij}|^2$ with an average, $\overline{|\gamma|^2}$,
\begin{eqnarray}
\Tr\left[\Delta M^p \Delta M^{n-p}\right] &=& \overline{|\gamma|^2}\left[\begin{array}{l}\sum_i{\left(|s_i|^2\left(1-|s_i|^2\right)^p\right)} \sum_j{\left(|s_j|^2\left(1-|s_j|^2\right)^{n-p}\right)} \\
	- \sum_k{\left(|s_k|^4\left(1-|s_k|^2\right)^n\right)}\end{array}\right] \nonumber \\
&=& \overline{|\gamma|^2}\left[\Tr\left[\rho(\Id-\rho)^p\right] \Tr\left[\rho(\Id-\rho)^{n-p}\right] - \Tr\left[\rho^2(\Id-\rho)^n\right]\right]
\end{eqnarray}
Inserting this expression into Eq. \ref{eqdHexplicit} yields
\begin{equation}
\dH \approx -\frac{\overline{|\gamma|^2}}{2}\sum\limits_{n=0}^{\infty}\frac{1}{n+1}\sum\limits_{p=0}^{n} \left[\Tr\left[\rho(\Id-\rho)^p\right] \Tr\left[\rho(\Id-\rho)^{n-p}\right] - \Tr\left[\rho^2(\Id-\rho)^n\right]\right].
\end{equation}

Finally, we can simplify this expression slightly by (1) taking advantage of the identity $\sum\limits_{n=0}^{\infty}{(\Id-\rho)^n} = \rho^{-1}$, and (2) rearranging the summation variables.
\begin{eqnarray}
\dH &\approx& -\frac{\overline{|\gamma|^2}}{2}\left[\sum\limits_{n=0}^{\infty}{\sum\limits_{p=0}^{n}{\frac{\Tr\left[\rho(\Id-\rho)^p\right] \Tr\left[\rho(\Id-\rho)^{n-p}\right]}{n+1}}} - \sum\limits_{n=0}^{\infty}{\Tr\left[\rho^2(\Id-\rho)^n\right]}\right] \\
&=& -\frac{\overline{|\gamma|^2}}{2}\left[\sum\limits_{n=0}^{\infty}{\sum\limits_{p=0}^{\infty}{\frac{\Tr\left[\rho(\Id-\rho)^p\right] \Tr\left[\rho(\Id-\rho)^{n}\right]}{n+p+1}}} - \Tr\left[\rho^2\sum\limits_{n=0}^{\infty}{(\Id-\rho)^n}\right]\right] \\
&=& -\frac{\overline{|\gamma|^2}}{2}\left[\sum\limits_{n=0}^{\infty}{\sum\limits_{p=0}^{\infty}{\frac{\Tr\left[\rho(\Id-\rho)^p\right] \Tr\left[\rho(\Id-\rho)^{n}\right]}{n+p+1}}} - 1\right] \label{eqdHfinal} 
\end{eqnarray}
Equation \ref{eqdHfinal} is the simplest form we have been able to achieve, except in very special cases, for $\Hh(\rho+\Delta)-\Hh(\rho)$.
\widetextend

\section{Probability distributions for additive decoherence factors} \label{appDFactors}

If $\ket{\psi}$ and $\ket{\psi'}$ are selected at random from the uniform ensemble of $\de$-dimensional quantum states, then the probability that $|\braket{\psi}{\psi'}| = \gamma$ (for $\gamma \in [0\ldots 1]$) is
\begin{equation}
p(\gamma) = 2(\de-1)\gamma(1-\gamma^2)^{\de-2}
\end{equation}
The additive decoherence factor $d$ is given by $d = -\log(\gamma)$, so that $\gamma = e^{-d}$ and $d \in [0\ldots \infty]$.  The probability distribution transforms as
\begin{eqnarray}
p(d) \mathrm{d}\!d &=& p(\gamma) \mathrm{d}\!\gamma \nonumber \\
p(d) &=& p(\gamma)\left|\frac{\mathrm{d}\!\gamma}{\mathrm{d}\!d}\right| \nonumber \\
     &=& e^{-d}p(\gamma) \nonumber \\
     &=& 2(\de-1)e^{-2d}\left(1-e^{-2d}\right)^{\de-2}
\end{eqnarray}

The decoherence factor for a collection of subenvironments is simply the sum of $d^{(i)}$ over the contributing subenvironments.  Ideally, we could obtain exact distributions $p_m(d)$ for a sum of $m$ such $d$-factors.  For an environment composed of qubits ($\de=2$), $p(d)$ is a 1st-order Poisson distribution, so $p_m(d)$ is just the $m$th order Poisson distribution (for details, see \cite{RBK04}).

For larger subenvironments ($\de > 2$), no such simple description exists.  However, the distribution functions $p(d)$ are well-approximated by Gaussian distributions.  We can treat the summing problem as a biased random walk, where the addition of another subenvironment represents a step forward with an approximately Gaussian-distributed stepsize.  

To compute the mean and variance of an $m$-step random walk, we first compute the mean value $\overline{d}$ and variance $\Delta d = \sqrt{\left(\overline{d^2}-\overline{d}^2\right)}$ for a single subenvironment.  Extrapolating to a collection of $m$ systems requires setting $\overline{d_m} = m\overline{d}$ and $\Delta d_m = \sqrt{m}\Delta d$.

For a single subenvironment, the mean $\overline{d}$ is given by $\overline{d} = \int_0^{\infty}{ d p(d) \mathrm{d}\!d}$.  This integral is somewhat nontrivial, involving an expansion in binomial coefficients:
\begin{eqnarray}
\overline{d} &=& 2(\de-1)\int_0^{\infty}{ d e^{-2d}\left(1-e^{-2d}\right)^{\de-2}\mathrm{d}\!d} \nonumber \\
&=& 2(1-\de)\int_0^{\infty}{ d e^{-2d} \sum_{k=0}^{\de-2}{\binom{\scriptstyle\de-2}{k} \left(e^{-2kd}\right)} \mathrm{d}\!d} \nonumber \\
&=& 2(1-\de)\sum_{k=0}^{\de-2}{\!(-)^k\binom{\scriptstyle\de-2}{k}\int_0^{\infty}{\!d \left(e^{-2(k+1)d}\right)\mathrm{d}\!d}}  \nonumber \\
&=& \frac{\de-1}{2}\sum_{k=0}^{\de-2}{\frac{(-)^k(\de-2)!}{(k+1)^2 k! (\de-2-k)!}} \nonumber \\
&=& \frac12\left(\Psi(\de)+\EMConst\right)
\end{eqnarray}
where $\Psi(\de)$ is the digamma function, and $\EMConst=0.5772\ldots$ is the Euler-Mascheroni constant.  A virtually identical calculation for $\overline{d^2}$ yields
\begin{equation}
\Delta{d}^2 = \frac{\pi^2}{24} - \frac{\Psi_1(\de)}{4}
\end{equation}
in terms of the trigamma function $\Psi_1(\de)$.  

\bibliographystyle{apsrev}
\bibliography{/home/rbk/bib/decoherence,/home/rbk/bib/Zurek,/home/rbk/bib/quantum,/home/rbk/bib/RBK,/home/rbk/bib/math}

\end{document}